# DNA Methylation Profiling in Melanoma: From Lesion Classification to Therapeutic Stratification

Jana T. Winterstein[a,b*], Lukas Heinlein[a,b*], Günter Raddatz[c], Carina Nogueira Garcia[a], Sarah Haggenmüller[a], Christoph Wies[a,b], Lucas Schneider[a], Annemarie Hoffsommer[a,b], Tim J. Zeuner[a,b], Friedegund Meier[d], Sarah Hobelsberger[d], Frank F. Gellrich[d], Mildred Sergon[d], Axel Hauschild[e], Lucie Heinzerling[f,g], Justin G. Schlager[f], Kamran Ghoreschi[h], Max Schlaak[h,i], Franz J. Hilke[h], Carola Berking[g], Markus V. Heppt[g,j], Michael Erdmann[g], Sebastian Haferkamp[k], Konstantin Drexler[k], Dirk Schadendorf[l], Wiebke Sondermann[l,m,n], Matthias Goebeler[o], Bastian Schilling[p], Daniel B. Lipka[q], Stefan Fröhling[q,r,s t], Felix Sahm[u], Jakob N. Kather[v,w,x,y], Yuri Tolkach[z], Jochen S. Utikal[aa], Benjamin Izar[bb], Yevgeniy R. Semenov[cc], Titus J. Brinker[a†]

a. Division of Digital Prevention, Diagnostics and Therapy Guidance, German Cancer Research Center (DKFZ), INF 223, 69120 Heidelberg, Germany
b. Medical Faculty, University Heidelberg, Heidelberg, Germany
c. Department of Epigenetics, German Cancer Research Center (DKFZ), Heidelberg, Germany
d. Department of Dermatology, Faculty of Medicine and University Hospital Carl Gustav Carus, Technische Universität Dresden, Germany, Skin Cancer Center at the University Cancer Center and National Center for Tumor Diseases Dresden, Dresden, Germany
e. Department of Dermatology, University Hospital (UKSH), Kiel, Germany
f. Department of Dermatology and Allergy, University Hospital, LMU Munich, Munich, Germany
g. Department of Dermatology, Uniklinikum Erlangen, Friedrich-Alexander-Universität Erlangen-Nürnberg; Comprehensive Cancer Center Erlangen – European Metropolitan Region Nürnberg, CCC Alliance WERA, Erlangen; Bavarian Cancer Research Center (BZKF), Erlangen, Germany
h. Department of Dermatology, Venereology and Allergology, Charité – Universitätsmedizin Berlin, Corporate member of Freie Universität Berlin and Humboldt-Universität zu Berlin, Berlin, Germany
i. Vivantes Hospital Spandau, Department of Dermatology and Allergology, Berlin, Germany
j. Dermpath München, Laboratory for Dermatopathology, Oral Pathology and Molecular Pathology, Munich, Germany
k. Department of Dermatology, University Hospital Regensburg, Regensburg, Germany
l. Department of Dermatology, Venereology and Allergology, University Hospital Essen, Essen, Germany
m. Department of Dermatology, Uniklinikum Erlangen, Friedrich-Alexander University Erlangen-Nürnberg, Erlangen, Germany.
n. Deutsches Zentrum Immuntherapie (DZI), Uniklinikum Erlangen, Erlangen, Germany.

o. Department of Dermatology, Venereology and Allergology, University Hospital Würzburg, National Center for Tumor Diseases (NCT) WERA Würzburg and Bavarian Cancer Research Center (BZKF), Würzburg, Germany
p. Goethe-University Frankfurt, University Hospital, Department of Dermatology, Germany.
q. Division of Translational Medical Oncology, German Cancer Research Center (DKFZ), Heidelberg, Germany
r. National Center for Tumor Diseases (NCT), NCT Heidelberg, a partnership between DKFZ and Heidelberg University Hospital, Heidelberg, Germany
s. German Cancer Consortium (DKTK), Core Center Heidelberg, Heidelberg, Germany
t. Division of Translational Precision Medicine, Institute of Human Genetics, Heidelberg University, Heidelberg, Germany
u. Department of Neuropathology, Heidelberg University Hospital, Heidelberg, Germany; CCU Neuropathology, DKTK, DKFZ
v. Else Kroener Fresenius Center for Digital Health, Faculty of Medicine and University Hospital Carl Gustav Carus, TUD Dresden University of Technology, Dresden, Germany
w. Department of Medicine I, Faculty of Medicine and University Hospital Carl Gustav Carus, TUD Dresden University of Technology, Dresden, Germany
x. Medical Oncology, National Center for Tumor Diseases (NCT), University Hospital Heidelberg, Heidelberg, Germany
y. Pathology & Data Analytics, Leeds Institute of Medical Research at St James's, University of Leeds, Leeds, United Kingdom
z. Institute of Pathology, University Hospital Cologne, Medical Faculty, University of Cologne, Cologne, Germany
aa. Department of Dermatology, Venereology and Allergology, University Medical Center Mannheim, Ruprecht-Karl University of Heidelberg, Mannheim, Germany
bb. Department of Medicine, Division of Hematology/Oncology, Herbert Irving Comprehensive Cancer Center, Columbia University Irving Medical Center, New York, NY, USA
cc. Department of Dermatology, Massachusetts General Hospital, Boston, Massachusetts; Harvard Medical School, Boston, Massachusetts; Harvard Data Science Initiative, Harvard University, Boston, Massachusetts.

*Equal contributions; †Correspondence: Titus.Brinker@nct-heidelberg.de*

# Abstract

DNA methylation provides a stable record of cellular identity, capturing epigenetic programs that distinguish specialized cell states despite a shared genome. Because malignant transformation and tumour progression are accompanied by extensive epigenetic remodeling, we hypothesized that the methylome of melanocytic lesions contains biologically and clinically relevant information for both diagnosis and disease progression. In a cohort of 1,001 tissue samples prospectively collected across eight German university hospitals profiled using Illumina Infinium MethylationEPIC arrays, we compared machine-learning models based on selected Cytosine phosphate Guanine (CpG) methylation sites with models incorporating biology-guided features, including epigenetic age acceleration, cell type composition and copy-number variation burden. In an external test set, the best diagnostic classifier was CpG-based and distinguished melanocytic nevi, non-invasive melanoma and invasive melanoma with a macro-averaged area under the receiver operating characteristic curve of 0.919 (95% CI: 0.878 to 0.952). Notably, across CpGs most strongly hyper- and hypomethylated between NV and IM, NIM showed an intermediate methylation profile, providing a molecular correlate of its diagnostic complexity. The best model for clinically relevant treatment group prediction, with AJCC stages grouped according to guideline-based management recommendations, relied on biology-guided features and achieved a macro-averaged mean absolute error of 0.627 (95% CI: 0.477 to 0.808). Together, these findings demonstrate that methylation-based models can capture both diagnostic identity and clinically relevant disease stratification, supporting DNA methylation as a promising biomarker for further validation and potential clinical translation.

# Main

Cellular identity and disease state are encoded not only by the genome but also by epigenetic regulation [1]. DNA methylation is a key component of this regulatory layer, capturing both cell type identity [2] and biologically distinct states within the melanoma lineage [3], thereby providing a promising molecular framework to resolve the spectrum of melanocytic lesions, from benign melanocytic nevi (NV) and non-invasive melanoma (NIM) to invasive melanoma (IM). As DNA methylation is remodelled during malignant transformation and tumour progression [4,5], the methylome may further capture clinically relevant features of disease progression that inform clinical staging. In melanoma this is particularly relevant, as the clinical stage e.g. defined by the American Joint Committee on Cancer (AJCC) informs both prognosis and treatment selection [6,7]. In clinical routine, melanoma diagnosis and staging is primarily established through histopathological assessment [6], yet evaluation of melanocytic lesions is subject to substantial inter-observer variability, particularly in borderline cases such as non-invasive melanoma (NIM) [8,9]. We hypothesized that the methylome of melanocytic lesions encodes molecular features that distinguish diagnostic categories and track clinically relevant disease progression.

Molecular data already offer promising complementary insights to enhance diagnosis and risk stratification: Driver-mutation analysis stratifies cutaneous melanoma into BRAF-, RAS-, NF1-, and triple-wildtype subtypes [10], and, together with immunohistochemical assays, informs both diagnosis and therapeutic decision-making [11,12]. Genome-wide DNA methylation profiling offers a complementary and increasingly powerful layer, as methylation patterns remain stable in formalin-fixed paraffin-embedded (FFPE) tissue, even with low input amounts or reduced sample quality [13]. As such, methylation-based classification has demonstrated potential across several cancer entities [14–16]. In melanocytic lesions specifically, methylation-based classifiers can distinguish melanocytic nevi (NV) from invasive melanoma (IM) with high accuracy [17]. However, NIMs have largely not been included in these approaches, despite representing diagnostically challenging and clinically highly relevant cases where additional computational support may be most beneficial [8].

To date, methylation profiling of melanocytic lesions has been used predominantly for diagnostic classification, leaving much of the underlying molecular information underexplored. Methylation array data enable inference of copy-number variations (CNVs) [18], deconvolution of cell type composition in a tissue [19], and quantification of epigenetic age acceleration (EAA) via methylation-based clocks [20]. Each of these readouts has independently been linked to cancer biology and outcome. CNV burden correlates with melanocytic malignancy and can be used as an additional diagnostic criterion for ambiguous lesions [21], tumour-infiltrating immune cell fractions show prognostic value in melanoma patients [22,23], and EAA has been shown to correlate with clinical outcome in several malignancies [24–26]. Building on this, features derived directly from methylation array data may offer potential to refine therapeutic assessment beyond diagnostic classification.

In this study, we present an integrative methylation-based framework for diagnosis and therapeutic stratification of melanocytic lesions. Our dataset comprises Illumina Infinium MethylationEPIC data for a cohort of 1,001 samples spanning NVs, NIMs, and IMs. Within this cohort, we systematically assessed how methylation array data can be leveraged for machine learning models by comparing data-driven Cytosine phosphate Guanine (CpG) methylation sites selection with the extraction of biologically interpretable methylation-derived features. We show that our methylation signature might support clinical decision making by informing diagnosis of melanocytic lesions and optimizing subsequent therapy selection.

# Results

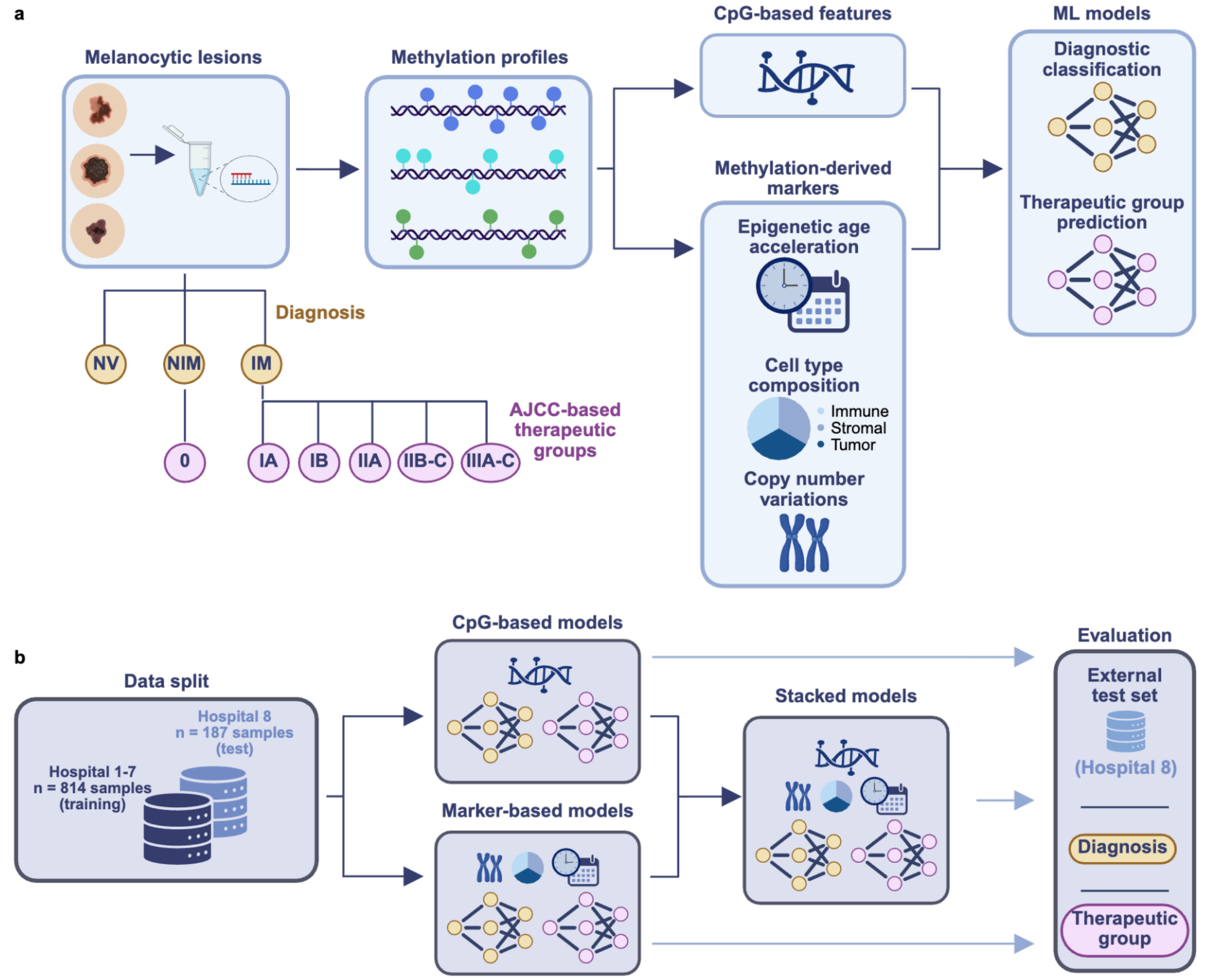

**Figure 1. Study design and workflow for methylation-based melanocytic lesion classification and therapeutic stratification.** **a** Melanocytic lesions were profiled using DNA methylation arrays to generate genome-wide methylation profiles. Two feature representations were derived: CpG-based features comprising selected individual methylation sites and biology-guided methylation-derived features, including epigenetic age acceleration, cell type composition and copy-number variation burden. These features were used to train machine-learning models for diagnostic classification of melanocytic lesions and therapeutic stratification, with AJCC stages grouped according to similarities in guideline-based clinical management recommendations. **b** Data were split by hospital, with samples from hospital 8 reserved as an independent external test set. Two complementary machine-learning models were trained on the remaining cohort: a CpG-based model using selected individual methylation sites and a marker-based model using methylation-derived biological features. Their prediction outputs were subsequently combined in a stacked model. All models were evaluated on the external test set for both diagnostic classification and prediction of therapeutic groups. *Created in https://BioRender.com*

To determine how different representations of the methylome support stratification of melanocytic lesions, we established an analytical framework addressing two clinically relevant tasks (**Figure 1**): three-class diagnostic classification of invasive melanoma (IM), non-invasive melanoma (NIM), and melanocytic nevi (NV); and therapeutic group prediction to support treatment decision-making. For the latter task, AJCC 8th edition stages [7] associated with the same treatment recommendations were combined into a single group (**Supplementary Table 1**). Our dataset comprised methylation data from 1,001 human tissue samples of melanocytic lesions (**Figure 2**) obtained from 923 patients, prospectively and consecutively collected in eight German university hospitals. Biology-guided markers reflecting residual epigenetic age acceleration (EAA), cell type composition, and copy-number variation (CNV) burden were derived from the methylation profiles. For each task we developed three models: one based on CpG features, one based on methylation-derived markers, and a stacked model that combined the predictions of the two base models (**Figure 1**). The stacked model was trained using their out-of-fold predictions to assess whether the two methylome representations provided complementary information. Performance was evaluated on the external test set.

Analysis of differentially methylated regions (DMRs) across the three diagnostic categories revealed the greatest number of high-effect DMRs between IM and NV (n = 2,267), followed by IM and NIM (n = 1,146) and NIM and NV (n = 193; **Supplementary Figure 3a, b**). When focusing on CpGs that were differentially methylated between IM and benign NV, NIM showed an intermediate position for hyper- and hypomethylated CpG groups (**Supplementary Figure 3 c, d**). The overlapping range of NIM with both NV and IM further illustrates the molecular continuum between these entities and may contribute to the diagnostic challenges associated with NIM.

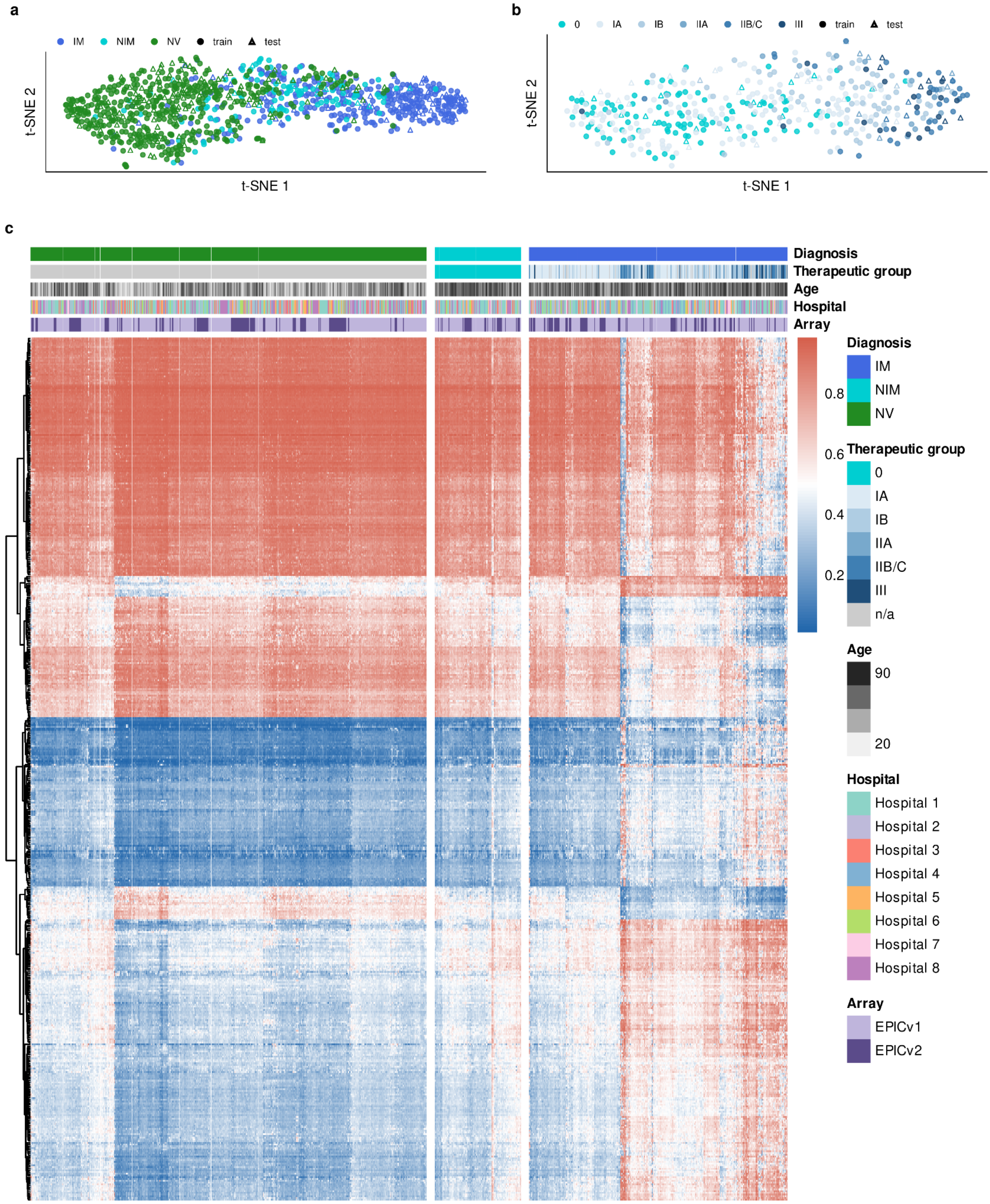

**Figure 2. Cohort overview showing the 500 CpGs most strongly associated with diagnosis**. CpGs were ranked by a per-CpG Kruskal-Wallis test against diagnosis on the training set; all panels show the complete cohort (n = 1,001; IM 349, NIM 117, NV 535). t-SNE embedding of the samples over these 500 CpGs, by **a** diagnosis and **b** therapeutic group. The therapeutic group panel is restricted to melanomas with an assignable AJCC stadium (n = 431; stage 0 n = 117, IA n = 154, IB n = 78, IIA n = 21, IIB/C n = 36,

III n = 25). **c** Heatmap of beta values of the 500 CpGs (rows) for cohort samples (columns). IM: invasive melanoma, NIM: non-invasive melanoma, NV: melanocytic nevus.

## Diagnostic Classification of Melanocytic Lesions

For the three-class diagnostic classification (IM, NIM, NV), models for each data modality (CpG-based, marker-based, stacked) were trained on the training set (Hospitals 1 to 7, n = 814; **Supplementary Table 2**), and then evaluated on the external test set (Hospital 8, n = 187). Results of the patient-level 5-fold cross-validation (CV) benchmark are available in **Supplementary Data 1**.

### Diagnostic Classification Results

The final CpG-based diagnostic classification model demonstrated strong overall performance (**Table 1** and **Figure 3**, confusion matrices are provided in **Supplementary Figure 1**), achieving a macro-averaged AUROC of 0.919 (95% CI: 0.878 to 0.952), thereby numerically outperforming models based on dermoscopic images or histopathological images on the same three-class task [27]. Class-wise evaluation revealed the highest discriminative performance for NV (AUROC: 0.946, 95% CI: 0.909 to 0.974), followed by IM (AUROC: 0.926, 95% CI: 0.878 to 0.964), whereas performance for NIM was lower (AUROC: 0.886, 95% CI: 0.809 to 0.947), indicating greater inherent diagnostic ambiguity for this class.

The final marker-based diagnostic classification model showed comparatively worse performance, achieving an AUROC of 0.843 (95% CI: 0.780 to 0.895). Class-wise evaluation demonstrated similar trends, with the highest discriminative performance for NV (AUROC: 0.900, 95% CI: 0.848 to 0.945), followed by IM (AUROC: 0.888, 95% CI: 0.821 to 0.944), whereas performance for NIM was also lower (AUROC: 0.742, 95% CI: 0.628 to 0.855).

The CpG-based model significantly outperformed the marker-based model on the primary endpoint (AUROC; paired patient-clustered bootstrap change in AUROC, ΔAUROC = 0.076, 95% CI: 0.033 to 0.119, $p < 0.001$; forest plots provided in **Supplementary Figure 2**). This was primarily due to the NIM class (DeLong, ΔAUROC = 0.143, Holm-adjusted $p = 0.007$), whereas the smaller CpG-based advantages for IM (DeLong, ΔAUROC = 0.038, $p = 0.048$, Holm-adjusted $p = 0.071$) and NV (DeLong, ΔAUROC = 0.046, $p = 0.035$, Holm-adjusted $p = 0.071$) were nominally significant but did not survive correction for multiple comparisons. Together, these results show that the CpG-based model reaches a level of diagnostic accuracy relevant for clinical use, particularly for the discrimination of NV and IM, while NIM remains the limiting class.

### Complementarity of the Diagnostic Classification Models

Of 187 external test cases, both models correctly classified 129 cases (69.0%), and were wrong in 25 cases (13.4%). The CpG-based model was correct in 23 cases (12.3%; one IM, eight NIM, 14 NV) in which the marker-based model was wrong, while there were ten cases (5.3%; three IM, one NIM, six NV) in which the CpG-based model was wrong and the marker-based model correct.

Stacking the CpG-based and marker-based predictions with a meta-classifier did not improve overall performance, consistent with the dominance of the CpG-based model among the discordant cases. The best-performing stacked model achieved an AUROC of 0.899 (95% CI: 0.841 to 0.947), numerically slightly below the CpG-based model alone (0.919), but not statistically significant (paired patient-clustered bootstrap, ΔAUROC = 0.020, 95% CI: -0.005 to 0.054, Holm-adjusted p = 0.124). Stacking did, however, significantly outperform the marker-based model (paired patient-clustered bootstrap, ΔAUROC = 0.055, 95% CI: 0.015 to 0.097, Holm-adjusted p = 0.020; per-class significant only for NV, DeLong, ΔAUROC = 0.052, Holm-adjusted p = 0.016). Class-wise performance of the stacked model followed similar trends: performance was highest for NV (AUROC: 0.952, 95% CI: 0.922 to 0.978), followed by IM (AUROC: 0.927, 95% CI: 0.883 to 0.963), and lowest for NIMs (AUROC: 0.817, 95% CI: 0.663 to 0.941).

**Table 1. Performance of the diagnostic classifiers.**

The best-performing metric in each category is highlighted in bold. AUROC: one vs. rest, macro-averaged area under the receiver operating characteristic curve, F1: macro averaged F1-score, CI: 95% confidence interval based on 1000 bootstrap iterations.

| **Metric (95% CIs)** | **AUROC ↑** | **Balanced Accuracy ↑** | **F1 ↑** |
|---|---|---|---|
| CpG-based | **0.919** (0.878-0.952) | **0.749** (0.652-0.838) | **0.712** (0.620-0.791) |
| Marker-based | 0.843 (0.780-0.895) | 0.595 (0.517-0.680) | 0.574 (0.494-0.654) |
| Stacked | 0.899 (0.841-0.947) | 0.739 (0.637-0.830) | 0.695 (0.609-0.771) |

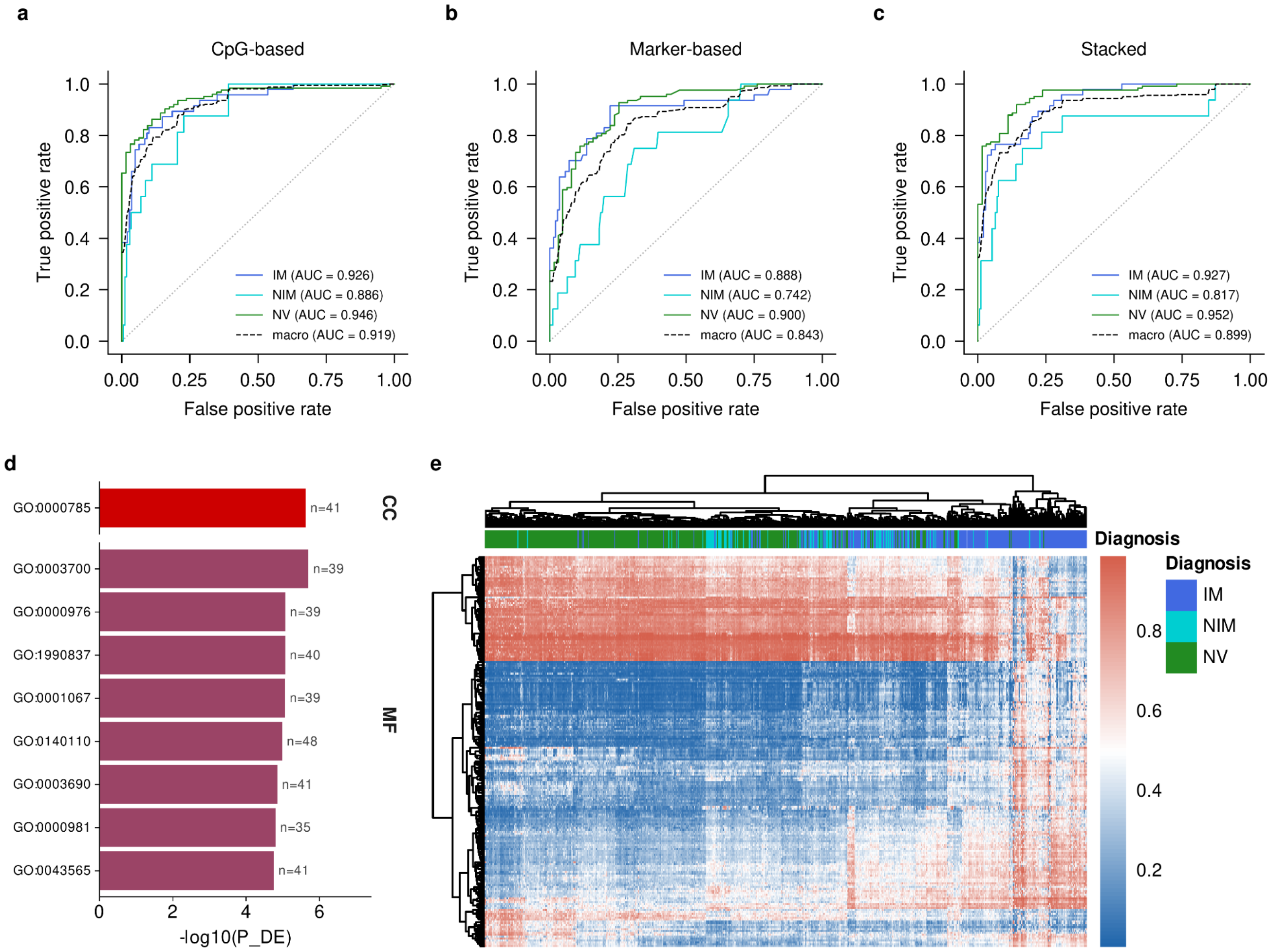

**Figure 3. Performance (top) and molecular characterization (bottom) of the best-performing diagnostic classification models**. ROC curves for invasive melanomas (IM, blue), non-invasive melanoma (NIM, cyan), melanocytic nevi (NV, green), and macro-averaged ROC curve (black dashed) for the model based on **a** CpGs, **b** methylation-derived markers, and **c** stacked. **d** Top enriched GO terms in the predictive selected CpG subset. Visualisations were restricted to terms with FDR < 0.05 and at least five contributing classifier CpG-hit genes. CC: cellular components, MF: molecular functions, GO:0000785: chromatin, GO:0003700: DNA-binding transcription factor activity, GO:0000976: transcription cis-regulatory region binding, GO:1990837: sequence-specific double-stranded DNA binding, GO:0001067: transcription regulatory region nucleic acid binding, GO:0140110: transcription regulator activity, GO:0003690: double-stranded DNA binding, GO:0000981: DNA-binding transcription factor activity, RNA polymerase II-specific, GO:0043565: sequence-specific DNA binding. **e** Heatmap of beta values of CpGs used in the CpG-based diagnostic classification model in significant high-effect DMRs (n = 205). Rows are CpGs, columns are samples and colored according to the diagnosis. IM: invasive melanoma, NIM: non-invasive melanoma, NV: melanocytic nevus.

### Feature Importance of the Diagnostic Classification Models

Gene ontology enrichment analysis of genes annotated to the CpGs most predictive of diagnosis returned 9 significant terms (FDR < 0.05), dominated by molecular function categories related to transcriptional regulation and DNA-binding (top terms: DNA-binding transcription factor activity [FDR = 0.025], chromatin [FDR = 0.025]; **Figure 3d**, **Supplementary Data 2**).

205 CpGs selected for diagnostic classification mapped to high-effect DMRs. Heatmap visualization of these regions showed no distinct NIM methylation pattern, but rather diffuse profiles overlapping both NV- and IM-associated signatures (**Figure 3e**).

Permutation importance for the marker-based model identified genome-wide CNV burden as the dominant predictor ($\Delta$AUROC = 0.20 ± 0.03; fraction of genome altered (FGA) and total CNV burden permuted jointly), ahead of the EpiScore cell type deconvolution estimates (0.12 ± 0.03), among which the $CD8^+$ T cell and melanocyte fractions ranked highest (0.029 ± 0.011 and 0.028 ± 0.015, respectively); Residual EAA contributed negligibly (0.014 ± 0.008, **Supplementary Figure 4a and b**). Estimated cell type composition across diagnostic classes is shown in **Supplementary Figure 5a**.

To complement the model-based analysis, we examined how each marker tracked diagnosis univariately on the training split (Kruskal-Wallis and Holm-adjusted post-hoc Mann-Whitney U tests for independent samples between diagnosis groups; **Figure 4**, **Supplementary Data 3**). All methylation-derived markers except the regulatory T cell fraction were significantly associated with diagnosis after correction, with the largest effects for total CNV burden and the $CD8^+$ T cell fraction, followed by FGA. Notably, this training-set univariate ordering is consistent with the test set permutation-importance ranking above.

For the stacked model, permuting the CpG base learner's class-probability outputs reduced AUROC by 0.27 ± 0.03, compared with 0.009 ± 0.017 for the marker base learner, confirming at the feature level that the meta-classifier relied almost exclusively on the CpG-based predictions (**Supplementary Figure 4c**). This is consistent with the dominance of the CpG-based model among the discordant cases and the absence of any significant gain from stacking.

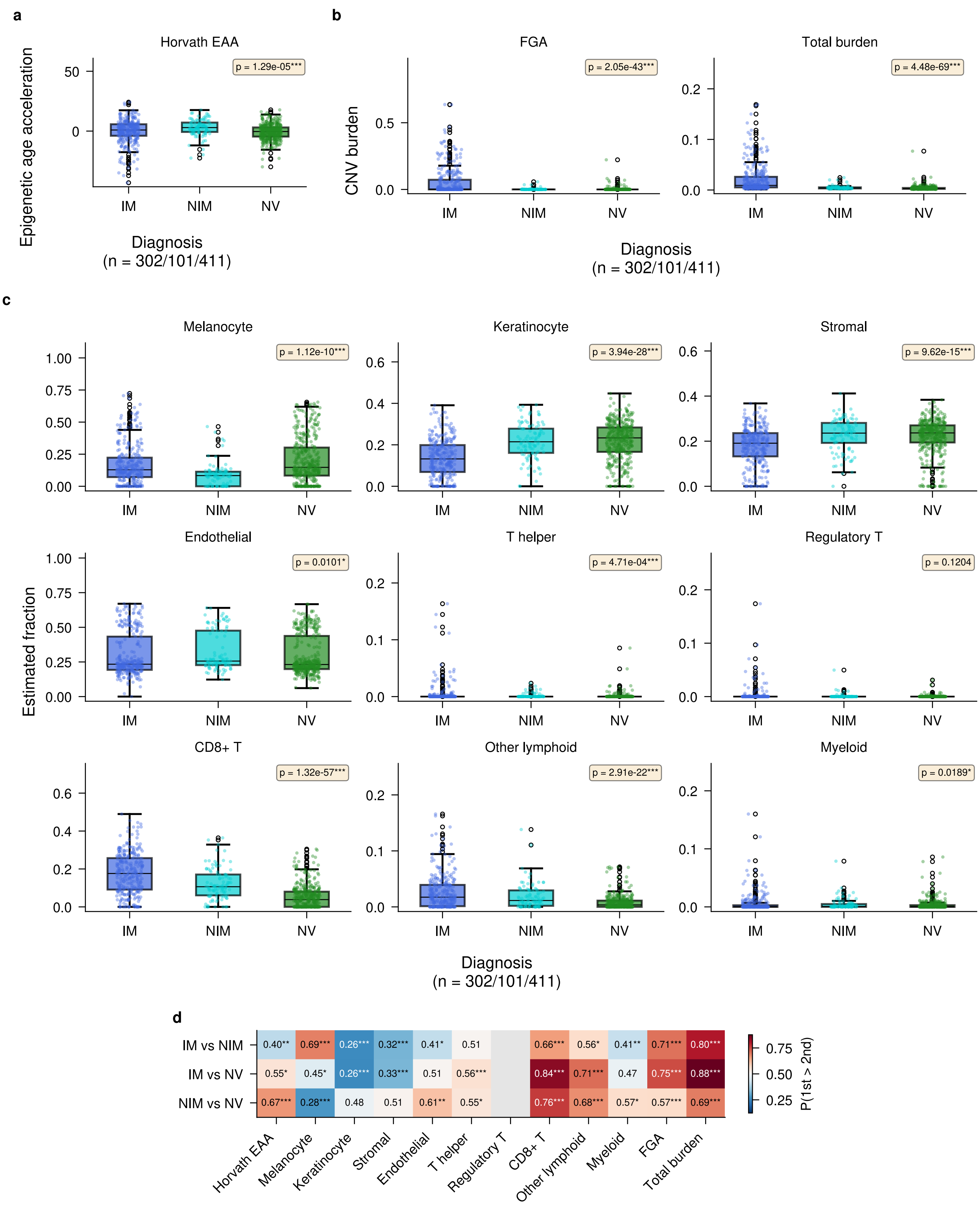

**Figure 4. Distribution of methylation-derived markers over diagnosis on the training set.** Distributions are shown for **a** Horvath residual epigenetic age acceleration (EAA), **b** copy-number variation (CNV) burden, **c** cell type composition. Holm-adjusted p-values of the Kruskal-Wallis omnibus test are displayed for each marker. **d** Association of each marker with each diagnostic contrast. Heatmap cells report Mann-Whitney $U/(n_1 \cdot n_2)$ test results, i.e. the probability that a randomly drawn sample of the pair's first group

exceeds one of its second group; 0.5 (white) indicates no rank separation, values above 0.5 (red) that the first group is higher and below 0.5 (blue) that the second is. The post-hoc Mann-Whitney U tests were run only for markers whose Kruskal-Wallis omnibus test was significant; regulatory T cell fraction did not reach significance and is shown in grey. IM: invasive melanoma, NIM: non-invasive melanoma, NV: melanocytic nevus, FGA: fraction of genome altered, ***: $p < 0.001$, **: $p < 0.01$, *: $p < 0.05$.

# Therapeutic Group Prediction for Melanoma

Analogously, for the therapeutic group prediction (groups comprising AJCC 8th edition stages [7] with comparable clinical management recommendations, as defined by the stage-specific treatment recommendations in clinical guidelines [28–30], **Supplementary Table 1**) models for each data modality (CpG-based, marker-based, stacked) were trained on the training set (Hospitals 1 to 7, n = 373; **Supplementary Table 3**), and then evaluated on the external test set (Hospital 8, n = 58). Results of the patient-level 5-fold CV benchmark are available in **Supplementary Data 4**.

### Therapeutic Group Prediction Results

The final CpG-based therapeutic group prediction model demonstrated strong overall performance (**Table 2** and **Figure 5**, confusion matrices provided in **Supplementary Figure 6**), achieving a macro-averaged MAE of 0.670 (95% CI: 0.539 to 0.814). Per-group MAE evaluation revealed better performance for groups requiring less intensive treatment (with the exception of group IIB/C, **Figure 5a**).

The final marker-based therapeutic group prediction model showed comparatively slightly better performance, achieving a macro-averaged MAE of 0.627 (95% CI: 0.477 to 0.808). Per-group MAE evaluation revealed balanced performance across all therapeutic groups (**Figure 5b**).

Although the macro-averaged MAE of the marker-based model was numerically lower than that of the CpG-based model (0.627 vs. 0.670), this difference was not statistically significant (paired patient-clustered bootstrap, ΔMAE = 0.043, 95% CI: -0.141 to 0.224, p = 0.65; **Supplementary Figure 7**). Together, these results show that methylation-based models predict therapeutic groups to within less than one group on average, a level of accuracy relevant for molecular stratification.

### Complementarity of the Therapeutic Group Prediction Models

Of 58 external test cases, both models correctly estimated 16 cases (27.6%), and were both equally wrong in 24 cases (41.4%). Their predictions differed in closeness to the truth in only 18 cases (31.0%), split near-evenly between them (CpG-based closer in ten, marker-based in eight). Overall, the errors of the two views are largely shared rather than complementary, suggesting limited gains for a combined model.

Stacking the CpG-based and marker-based predictions with a meta-regressor did not improve performance. The best-performing stacked model achieved a macro-averaged MAE of 0.758 (95% CI: 0.603 to 0.947), numerically higher (worse) than both base models, but neither difference reached significance in the paired patient-clustered bootstrap (versus the CpG-based model: Δmacro-MAE = -0.087, 95% CI: -0.205 to 0.037, Holm-adjusted p = 0.28; versus the marker-based model: Δmacro-MAE = -0.130, 95% CI: -0.303 to 0.046, Holm-adjusted p = 0.28).

**Table 2. Performance of the methylation classifier.**

The best-performing metric in each category is highlighted in bold. MAE: macro-averaged Mean Absolute Error, QWK: Quadratic Weighted Kappa, CI: 95% confidence interval based on 1000 bootstrap iterations.

| **Metric (95% CIs)** | **MAE ↓** | **QWK ↑** | **Spearman ρ ↑** |
|---|---|---|---|
| CpG-based | 0.670 (0.539-0.814) | 0.831 (0.736-0.893) | **0.822** (0.714-0.890) |
| Marker-based | **0.627** (0.477-0.808) | **0.852** (0.769-0.911) | 0.801 (0.658-0.883) |
| Stacked | 0.758 (0.603-0.947) | 0.760 (0.646-0.837) | 0.798 (0.672-0.869) |

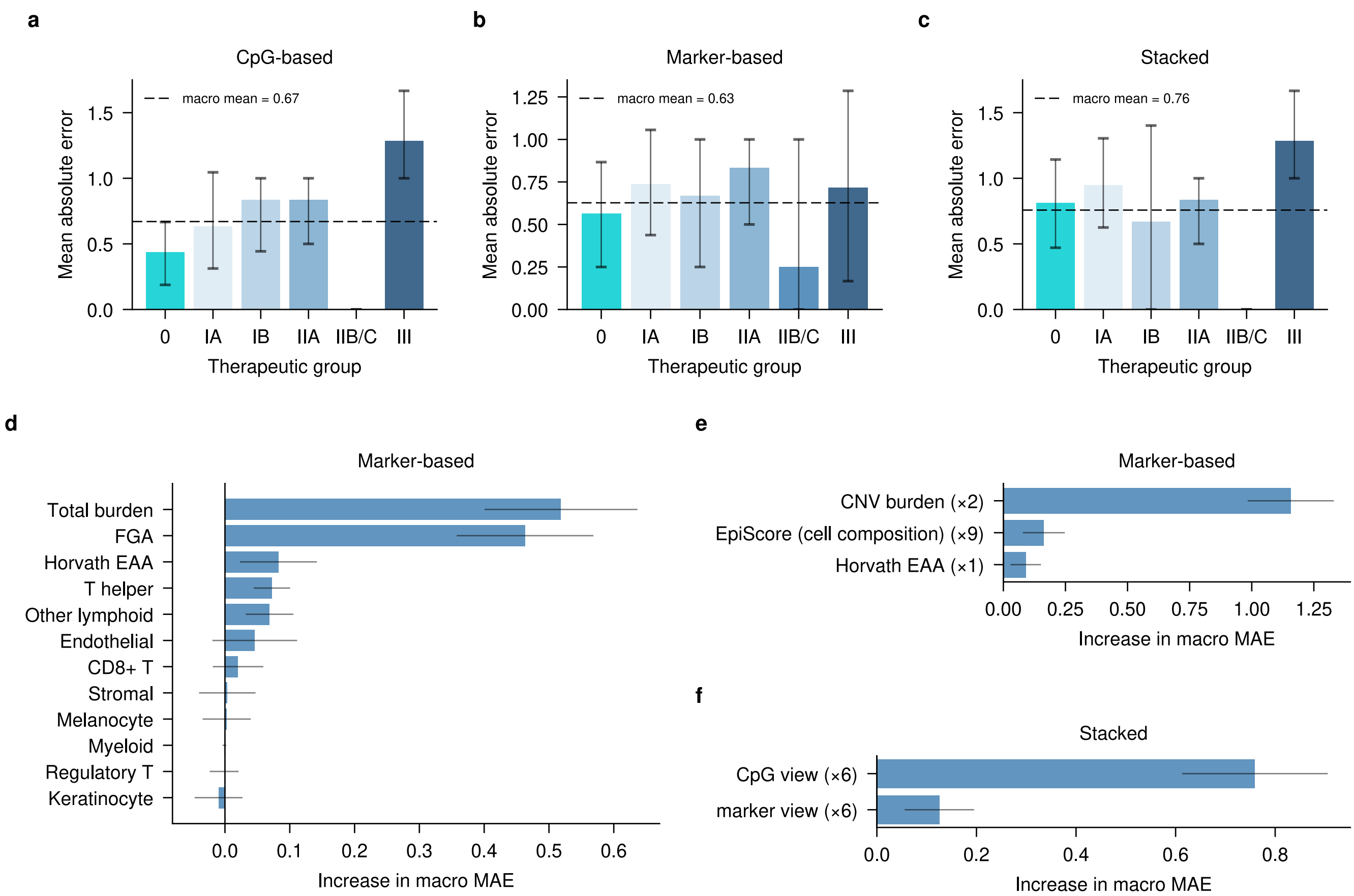

**Figure 5. Performance of the best-performing therapeutic group prediction models by means of the per-group MAE (top) and permutation importance for the marker-based and stacked model (bottom).** Results for the model based on **a** CpGs, **b** methylation-derived markers, and **c** stacked. Permutation importance for **d** all features of the marker-based model, **e** the grouped features of the marker-based model, and **f** the grouped features of the stacked model.

**Feature Importance of the Therapeutic Group Prediction Models**

Gene Ontology enrichment analysis of genes linked to the CpGs most predictive of therapeutic groups identified no significantly enriched terms at FDR < 0.05, indicating no detectable functional convergence among these genes (**Supplementary Data 5**).

Permutation importance for the marker-based model (**Figure 5d**, **e**) identified genome-wide CNV burden as the dominant predictor (Δmacro-MAE = 1.16 ± 0.17 therapeutic groups when FGA and total CNV burden were permuted jointly), ahead of the EpiScore cell type deconvolution estimates (0.16 ± 0.08) and EAA (0.09 ± 0.06). Within the EpiScore block the T helper and other lymphoid cell fractions ranked highest (0.072 ± 0.028 and 0.069 ± 0.037, respectively). Estimated cell type composition across therapeutic groups are shown in **Supplementary Figure 5b**.

Univariate analysis of the same methylation-derived markers against therapeutic group on the training split (Spearman ρ and Kruskal-Wallis, Holm-adjusted; **Figure 6**, **Supplementary Data 6**) showed that CNV correlated most strongly with therapeutic groups (total CNV burden ρ = +0.68; FGA ρ = +0.60; both Holm-adjusted $p < 0.001$), followed by the keratinocyte (ρ = -0.50, Holm-adjusted $p < 0.001$) and melanocyte (ρ = +0.44, Holm-adjusted $p < 0.001$) fractions. Residual EAA correlated only weakly (ρ = -0.17, Holm-adjusted $p = 0.004$), and the T helper, regulatory T cell, and other lymphoid cell fractions showed no significant monotonic trend with therapeutic groups. The dominance of CNV again matches its rank as the leading predictor in the marker-based model's permutation importance.

For the stacked model, permuting the CpG base learner's therapeutic group-probability outputs increased macro-MAE by 0.76 ± 0.15 therapeutic groups, versus 0.13 ± 0.07 for the marker base learner, showing that the meta-regressor relied far more heavily on the CpG-based predictions (**Figure 5f**). This tracks the base learners' cross-validated performance (**Supplementary Data 4**), where the CpG-based model (macro-MAE 0.79 ± 0.19) outperformed the marker-based model (0.94 ± 0.25) on the out-of-fold predictions used to fit the meta-regressor, even though the two performed comparably on the test set.

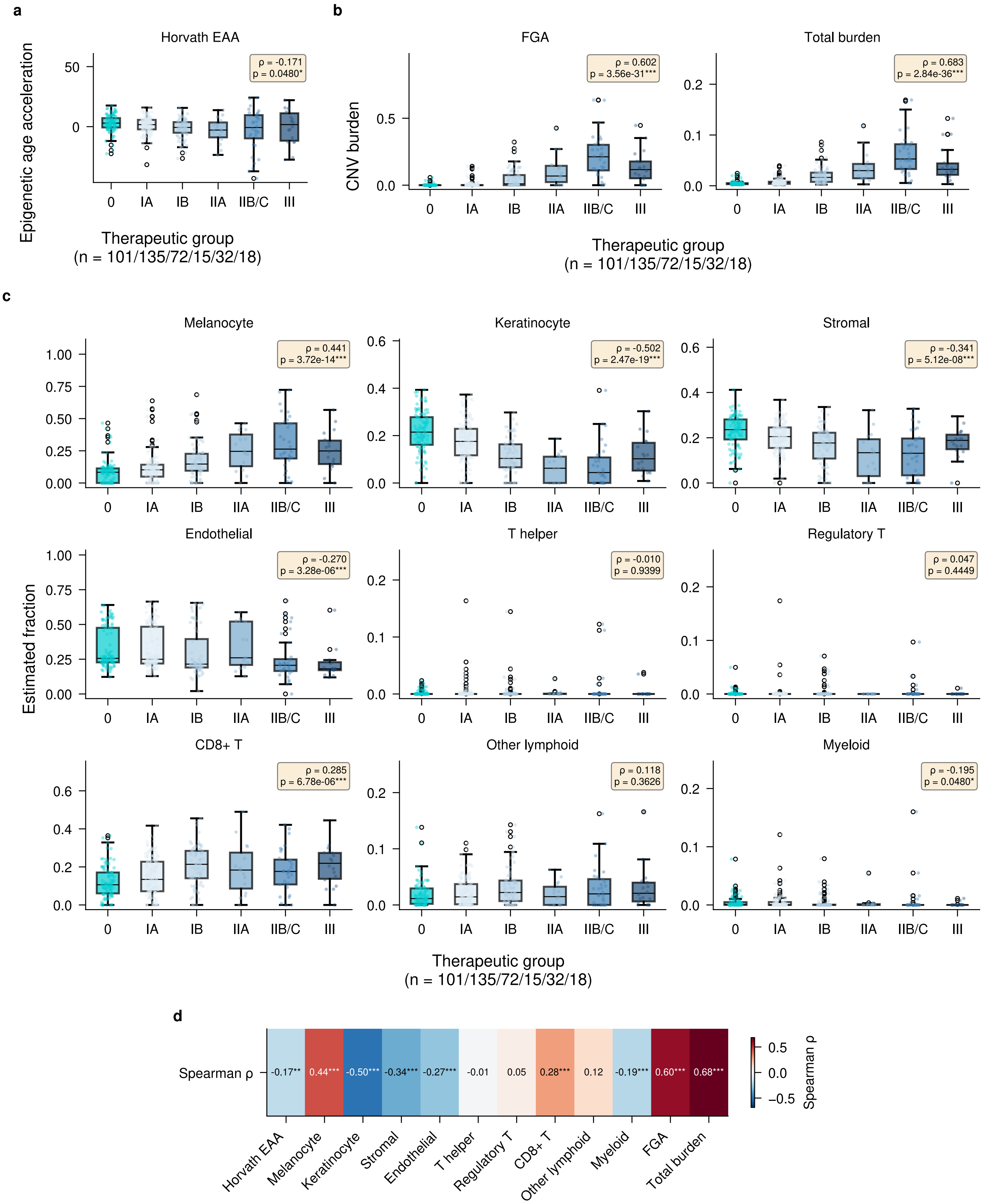

**Figure 6. Distribution of methylation-derived markers over therapeutic groups on the training set.** Distributions are shown for **a** Horvath residual epigenetic age acceleration (EAA) **b** copy number variation (CNV) burden **c** cell type composition. Holm-adjusted p-values of the Kruskal-Wallis omnibus test and Spearman's correlation ρ against the ordinal group code are displayed for each marker. **d** Overview of the correlation of all marker features with the therapeutic group. Table entries report Spearman's ρ; 0 (white)

indicates no monotonic association, values above 0 (red) that the marker increases with the therapeutic group and below 0 (blue) that it decreases. FGA: fraction of genome altered, ***: $p < 0.001$, **: $p < 0.01$, *: $p < 0.05$.

# Discussion

This study provides evidence that DNA methylation profiling can resolve clinically relevant categories of melanocytic lesions and capture molecular features associated with melanoma stratification. In a prospective cohort of 1,001 human tissue samples, methylation-based classifiers achieved strong overall performance in distinguishing NV, NIM, and IM, with highest discriminative performance for NV and IM and reduced performance for NIM. For therapeutic stratification within melanoma, methylation-based models achieved a mean prediction error of less than one group for the assignment across the six pooled AJCC-based groups.

Our findings extend prior work by Conway et al., who developed a 40-CpG methylation classifier to distinguish primary melanoma from NV [17]. Comparison of the CpG positions selected for our diagnostic classifier with the DAVID-based GO analysis reported for the Conway et al. melanoma classifier showed thematic convergence on transcriptional-regulatory and DNA-binding categories [17], pointing to a potential shared theme of dysregulated transcriptional control in melanocytic neoplasia.

In contrast to the binary framework of Conway et al. [17], our study addressed a three-class classification problem, including the clinically relevant but diagnostically challenging non-invasive category. The reduced classification accuracy for NIM observed in our study is consistent with the recognized diagnostic difficulty of this category, as expert pathologist agreement is substantially lower for NIM than for NV or IM [8]. Together with genomic evidence that NIM represent an early progression state, characterized by emerging TERT promoter mutations and occasional copy-number alterations, while lacking later events such as biallelic CDKN2A loss that arise with invasion [31], this suggests that NIM may occupy a biologically and diagnostically intermediate position between benign melanocytic lesions and invasive disease. This interpretation is supported by our findings, as NIM samples did not form a distinct methylation cluster based on classifier-selected CpGs but instead showed diffuse patterns overlapping with both NV- and IM-associated signatures. Across the 20,000 CpGs most strongly hyper- and hypomethylated between NV and IM, the median NIM lesion was positioned near the midpoint between the two groups, while the overall NIM range overlapped substantially with both.

In addition to data-driven CpG selection, we used prior knowledge-guided processing of the same methylation array data to derive biologically interpretable features, including residual EAA, methylation-derived cell type composition, and CNV burden. Similar informed machine learning approaches have been applied across biomedical settings by integrating biological information into feature aggregation or model architecture, thereby improving interpretability of high-dimensional omics-based prediction models [32–34], which is particularly important in light of scenarios where false reliance may occur (e.g. in a clinical setting) [35]. In our case, the resulting performance patterns indicate task-specific value: CpG-level signatures were most informative for diagnostic classification, whereas methylation-derived markers contributed more strongly to therapeutic stratification.

This adds to existing work showing that integrated molecular profiling can reflect broader tumour biology beyond malignancy classification. Griewank et al. demonstrated that combined mutation analysis, copy-number profiling, and DNA methylation profiling can distinguish cutaneous melanoma metastases from other melanocytic tumours of the central nervous system [36]. Similarly, methylation profiling across melanoma subtypes separated uveal melanomas from other primary sites, while identifying site-specific differences in promoter methylation, copy-number alterations, and mutations [37]. Our study extends this concept by demonstrating that methylation array data can be transformed into biologically interpretable features that improve therapeutic stratification within melanoma.

Across the therapeutic groups, total CNV burden and FGA showed the strongest positive correlations, consistent with previous associations of chromosomal instability and specific copy-number alterations with adverse outcome in melanoma patients [38]. Estimated melanocyte and $CD8^{+}$ T cell fractions were also positively correlated, whereas non-melanocyte and non-lymphoid cell fractions, as well as EAA, decreased across groups. The inverse association of EAA with the therapeutic group was consistent with its distribution across diagnostic categories, where residual EAA peaked in NIM and showed relative deceleration in IM. Although EAA has been associated with increased cancer risk across tissues [39], this pattern suggests that invasive lesions may not simply reflect continued epigenetic aging, but may acquire distinct clock profiles during progression. Consistent with this, Rodríguez-Paredes et al. similarly identified a subclass of non-melanoma skin tumours characterized by reduced DNA methylation age [40].

The accessibility of methylation profiling is expanding through minimally invasive sampling approaches, such as tape stripping [41,42], supporting the scalability for clinical implementation. In line with the proposed workflow for the integration of DNA methylation profiling into tumour diagnostics by Koelsche and von Deimling [43], translation of these approaches into routine would require implementation as a decision-support tool alongside histopathological evaluation, informing but not replacing (dermato-)pathologist assessment. This role is consistent with recent expert-informed dermoscopic and whole slide image classifiers, which achieved performance comparable to local dermatologists and pathologists while providing complementary diagnostic information [27]. Interestingly, our methylation-based classifier performed in a similar range as reported for these image-based models, and both approaches showed reduced performance for NIM, underscoring the diagnostic ambiguity of this category [27].

Several factors should be considered when interpreting these findings. First, diagnostic ground truth remains inherently imperfect, particularly for NIM, where histopathological assessment is subject to substantial interobserver variability. Second, AJCC 8th edition stage information could not always be assigned to a defined clinical time point, as recorded stages may reflect either the initial diagnostic assessment or later clinical updates. Third, therapeutic groups are based on guideline recommendations, rather than follow-up data on how patients were actually treated. Finally, class imbalance, especially for NIM and for some therapeutic groups, may have affected model training, class-wise performance estimates, and the stability of subgroup-specific predictions.

This study extends methylation profiling beyond diagnostic classification toward integrated molecular characterization relevant to therapeutic stratification in melanoma. A multi-layered framework incorporating EAA, tumour cell type composition, and CNV burden provides complementary information that may refine assessment of therapy selection. Nevertheless, the clinical value of methylation-based therapeutic group prediction beyond established histopathological staging requires validation in larger cohorts with extended follow-up.

# Methods

## Ethics Statement and Reporting Standards

Ethics approval was obtained from the ethics committees at the Technical University of Dresden (BO-EK-53012021), the Friedrich-Alexander University Erlangen-Nuremberg (69_21 Bc), the LMU Munich (21-0182), the University of Regensburg (20-2190-103), the University of Würzburg (293/20_z) and from the University Hospitals Mannheim (2010-318N-MA and 2014-835R-MA), and Essen (20-9784-BO). From the Charité Berlin a separate ethics approval was not required. For a study across multiple centres an additional approval is not necessary according to §15(2) of the Berufsordnung der Ärztekammer Berlin (professional code of conduct of the medical association Berlin) if there is approval from another ethics board of a German University or medical association. Patients provided informed written consent. The study was performed in accordance with the Declaration of Helsinki.

## Data collection

### Patient Cohort

Study participants were prospectively and consecutively enrolled at eight university hospitals in Germany (Berlin, Dresden, Erlangen, Essen, Mannheim, Munich, Regensburg, Würzburg) from April 2021 to February 2023 as part of the Skin Classification Project [8,44]. The dataset used in this study comprised 1,001 prospectively and consecutively collected lesions obtained from 923 patients, including IMs (n = 349), NIMs (n = 117) and NV (n = 535), diagnosed by at least one (dermato-)pathologist. For therapeutic stratification, melanoma samples with available AJCC 8th edition stage [7] were assigned to six ordinal groups according to stage-specific management recommendations [28–30]: group 0 (stage 0, n = 117), wide excision, with topical imiquimod or radiotherapy as possible alternatives; group IA (stage IA, n = 154), wide excision; group IB (stage IB, n = 78), wide excision and consideration/offering of sentinel lymph-node (SLN) biopsy; group IIA (stage IIA, n = 21), wide excision with SLN biopsy to be discussed and offered; group IIB/C (stages IIB-IIC, n = 36), wide excision and SLN biopsy, with consideration/offering of adjuvant systemic or regional therapy; and group III (stages IIIA-IIID, n = 25; no patients with stage IIID disease represented in this cohort), wide excision, selective completion lymph-node dissection, BRAF testing, and stage-dependent consideration or recommendation of adjuvant or neoadjuvant systemic therapy. Stage IV was excluded because of insufficient sample size.

For a detailed breakdown of the patient and lesion-specific characteristics please refer to **Table 3** and **Supplementary Table 4**.

**Table 3. Lesion Characteristics.** IM: invasive melanoma, NIM: non-invasive melanoma, NV: melanocytic nevus, std: standard deviation, IQR: interquartile range.

| **Lesion Characteristic** | | **n = 1,001** | **%** |
|---|---|---|---|
| Diagnosis | | | |
| | IM | 349 | 34.9 |
| | NIM | 117 | 11.7 |
| | NV | 535 | 53.5 |
| Location | | | |
| | Face/scalp/neck | 142 | 14.2 |
| | Palms/soles | 20 | 2.0 |
| | Upper extremities | 125 | 12.5 |
| | Lower extremities | 190 | 19.0 |
| | Back | 313 | 31.3 |
| | Abdomen | 103 | 10.3 |
| | Chest | 73 | 7.3 |
| | Buttock | 23 | 2.3 |
| | Genitalia | 9 | 0.9 |
| | Unknown | 3 | 0.3 |
| **Melanoma Characteristic** | | **n = 466** | |
| Breslow | [mean (std, IQR)] | 1.62 | 2.4 (1.4) |
| AJCC 8th Edition Stage | | | |
| | 0 | 117 | 25.1 |
| | IA<br>IB | 154<br>78 | 33.1<br>16.7 |
| | IIA<br>IIB<br>IIC | 21<br>23<br>13 | 4.5<br>4.9<br>2.8 |
| | IIIA<br>IIIB<br>IIIC | 4<br>6<br>15 | 0.9<br>1.3<br>3.2 |
| | IV | 1 | 0.2 |
| | Unknown | 34 | 7.3 |

**Tissue Processing**

From each FFPE tissue block, six serial 3-4 µm sections were cut and prepared on uncoated glass slides. One section was stained with haematoxylin and eosin, and tumour regions were annotated by a lead senior dermatopathologist (JSU). These annotations guided the selection and manual dissection of corresponding tumour areas from the remaining unstained serial sections, which were subsequently used for DNA extraction with the QIAamp DNA FFPE Advanced Kit (QIAGEN) according to the manufacturer's instructions. Epigenetic profiling was performed using the Illumina Infinium MethylationEPIC v1 and v2 BeadChip arrays.

## Data Preprocessing

DNA methylation at each CpG site is summarised as a beta value, the ratio of the methylated probe intensity to the intensity of all probes, which ranges from 0 (fully unmethylated) to 1 (fully methylated) and approximates the proportion of methylated alleles at that CpG site in the sample. Beta values were derived from raw Illumina IDAT files using the *sesame* package in R [45], following recommended preprocessing procedures for human EPIC arrays. This included the removal of poorly designed probes, correction for non-linear dye bias, p-value with out-of-band array hybridization (pOOBAH) filtering (retaining probes with detection $p<0.05$), and normal-exponential out-of-band normalization. Probes overlapping single nucleotide polymorphisms (SNPs), cross-hybridizing probes, and ctl-/rs-probes were excluded using the *DMRcate* package [46]. Samples with individual probe detection rates lower than 90 % at pOOBAH detection p = 0.05 were excluded from further analysis, resulting in n = 1,001 samples. To account for probe name changes between array versions, Illumina Infinium MethylationEPIC v1 and v2 BeadChip arrays were harmonized based on genomic coordinates (hg38) using the EPIC manifest described by Zhou et al. [47], retaining only probes with overlapping genomic positions. Probes located on the X and Y chromosomes, lacking a valid genomic mapping, or failing the pOOBAH detection threshold in more than 20% of samples were excluded. For genomic positions represented by multiple probes, beta values were averaged. Cases from hospitals 1 to 7 were used for training and validation purposes, while hospital 8 was held out for external testing (**Supplementary Tables 2** and **3**). After creating the data split, remaining missing values in each subset were imputed separately using the *MethylLImp2* package [48], incorporating chromosome information from the prior mapping.

## Methylation-derived Features

In addition to the CpG-level beta matrix, three classes of biologically interpretable per-sample features were derived. To avoid harmonization artefacts, all three were computed on an intermediate output of the preprocessing pipeline prior to cross-array harmonization, with EPIC v1 and EPIC v2 cohorts processed separately.

**Residual EAA**

EAA was estimated using Horvath's Skin & Blood clock [20], as implemented in the *methylclock* package [49]. Residual EAA was defined, as in *methylclock's ageAcc2*, as the residual of an ordinary least-squares regression of predicted DNAm age on chronological age. Because methylclock fits this regression across all samples supplied to it, each sample contributes to the regression yielding its own residual. We therefore refit the same model with a restricted fit population: residuals for the final model and the external test set derive from a line fit on the training cohort alone, and during CV each validation sample's residual derives from a line fit on the training rows of the fold in which it was held out.

**Cell type Composition**

To enable cell type deconvolution of bulk DNA methylation profiles, a custom tissue-specific DNAm reference matrix was constructed using *EpiSCORE* [19]. We leveraged the Sanger Human Skin Cell Atlas as the basis for reference imputation [50]. Prior to reference construction, cell type annotations were consolidated into 15 biologically relevant cell types based on relevance to cutaneous melanoma and NV, with a minimum threshold of 300 cells per group. Methylation distinguishability was assessed using pseudobulk Pearson correlation in expression space. With the exception of T cells, no subtypes were retained due to their distinct roles in tumour immune evasion. Beta-value matrices from Illumina Infinium MethylationEPIC arrays were summarized at the gene promoter level and cell type fractions were estimated using weighted robust partial correlation [51]. This yielded fractions for nine cell types (Melanocyte, Keratinocyte, Stromal, Endothelial, T helper cells, regulatory T cells, CD8+ T cells, other Lymphoid, Myeloid).

**CNV Burden**

CNV profiles were inferred from DNA methylation array data using the *conumee2.0* package [18]. The analysis was performed according to the recommended workflow, leveraging normalized intensity signals to estimate genome-wide copy number changes for each sample. EPIC v1 and EPIC v2 cohorts were processed separately against their respective genome builds (hg19 and hg38), and autosomal probes were retained for downstream analysis. To prevent data leakage from the external test cohort, only NV samples from the training Hospitals were used as the diploid reference panel; when a NV sample itself was queried, it was excluded from its own reference set in a leave-one-out fashion. For the CV experiments described below, the NV reference was additionally rebuilt from each training fold only, so validation samples never contributed to their own reference. For each sample, intensities were fit against the reference, binned, and segmented, after which segments with $|\log2 \text{ ratio}| > 0.1$ were classified as altered (gain if positive, loss if negative). Per-sample burden metrics included FGA and total CNV burden (length-weighted mean of $|\log2 \text{ ratio}|$ across all segments).

# Machine Learning

Two tasks were addressed within a unified 5-fold CV framework: (i) three-class diagnostic classification (IM / NIM / NV) and (ii) therapeutic group prediction based on AJCC 8th edition stages [7] grouped into ordinal categories from 0-5 (0 / IA / IB / IIA / IIB-C / IIIA-C, **Supplementary Table 1**). For each task, two feature sources were evaluated: raw CpGs, and methylation-derived markers. A third model was then stacked on top of these two, taking the out-of-fold predictions of the two base models as its only inputs. To ensure reproducibility, a fixed random seed was used throughout the whole project.

**CV and external evaluation**

All benchmarks used patient-stratified 5-fold CV on the training cohort, so that slides from the same patient never split across folds. Stratification used the target directly in both tasks: the diagnosis label for classification and the full ordinal therapeutic group code for staging. To address class imbalance, the Synthetic Minority Over-sampling Technique [52] was applied to the training portion of each fold. The test cohort comprised all samples from a single clinical site (hospital 8) held out in its entirety from CV and model selection, providing an independent, cross-site evaluation. Each selected model was retrained on the full training cohort and evaluated on this site only once.

**CpG-based Models**

Raw beta values were passed through a filter-reducer-classifier pipeline fit independently per fold. Filters comprised both data-driven and knowledge-driven filtering strategies. Data-driven filters included interquartile range, variance, differential methylation across classes (ANOVA F-test for classification, absolute Spearman correlation for ordinal regression), LASSO, ElasticNet, and Boruta. Knowledge-driven curation used melanoma-specific gene sets and ChromHMM-defined chromatin states. Chromatin states were annotated using the 15-state ChromHMM model from the Roadmap Epigenomics Project, based on the ChromHMM framework and corresponding reference epigenome annotations for melanocytes (E059) [53,54]. State pairs evaluated were repressed regions (E13/E14), active transcription sites (E1/E2), and enhancers (E6/E7).

Dimensionality reduction (applied after filtering when used) included principal component analysis (PCA; 100, 200, 300 components), partial least squares (PLS; 20, 30 components), undercomplete and variational autoencoders (64 latent dimension each), and aggregation of beta values by ChromHMM state (15 states).

Classification architectures comprised Random Forest, XGBoost, logistic regression, ElasticNet logistic regression, support vector machines (SVM), multilayer perceptrons (MLP), and 1D convolutional neural networks (CNN). For therapeutic group prediction, architectures comprised ordinal-regression models including Ridge, Lasso, ElasticNet, support vector regression (SVR), Random Forest regression, XGBoost regression, mord OrdinalRidge and LogisticAT [55], ordinal gradient boosting (ogboost) [56], and three neural ordinal regressors: CORAL [57], CORN [58], and a cumulative-logits MLP trained with binary cross-entropy.

**Marker-based models**

The methylation-derived features (Horvath residual EAA, nine EpiSCORE cell type fractions, and the two CNV burden metrics) were combined into a single low-dimensional tabular matrix and used as input to a compact model set comprising logistic regression, ElasticNet logistic regression, SVM, Random Forest, XGBoost, and a small MLP for classification, and the same set of ordinal-regression models for therapeutic group prediction. Scale-sensitive estimators (logistic regression, ElasticNet, SVM/SVR, and MLP) were wrapped in a StandardScaler pipeline fit on the training portion of each fold only, to account for the different scales (years, [0,1] fractions, log2 units). For each CV fold, the validation rows' CNV burden was replaced with values computed against the per-fold training-only NV reference (see CNV burden above), so no validation sample ever saw its own reference. Filter and reducer stages were bypassed for this feature source.

**Stacked models**

To assess complementarity between the raw CpG signal and the derived biological features, a stacked feature representation was constructed by combining the out-of-fold CV predictions of the best CpG-based model with those of the best marker-based model (two-base-learner stacking). Both views were placed on a common probability scale (per-class probabilities for diagnosis, and per-group probabilities across the six ordinal categories for therapeutic groups) yielding a compact meta-feature matrix (two times the number of classes). Because each base learner's predictions are themselves held-out CV predictions rather than in-sample fits, no validation sample contributed to either base-model output it was later scored against. The same compact model set as for the marker-based pipeline was then trained on this matrix under the same patient-stratified 5-fold CV.

**Benchmarking and model selection**

All combinations of feature source, filter, reducer, and model were evaluated by 5-fold CV. **Supplementary Data 1** and **4** report per-configuration mean and standard deviation across folds. For diagnostic classification, metrics were macro-averaged AUROC as the primary endpoint, balanced accuracy and macro F1 score as secondary endpoints. For therapeutic group prediction, ordinal metrics were used: macro-averaged MAE across therapeutic groups (primary endpoint), quadratic weighted Cohen's κ, and Spearman's ρ between predicted and true group. Based on these criteria, a final configuration was selected per feature source. For diagnostic classification these were the Boruta-filtered SVM without dimensionality reduction (CpG), a Random Forest (markers), and a MLP (stacked); for therapeutic group prediction, the Boruta-filtered ogboost without dimensionality reduction (CpG), ogboost (markers), and mord OrdinalRidge (stacked). Each selected pipeline was retrained on their respective full training set (CpGs, markers, out-of-fold predictions) and then finally evaluated on the external test set.

## Interpretability

### Gene set enrichment

The selected CpGs of each CpG-based model were tested for Gene Ontology enrichment using the *missMethyl* package for R [59], correcting for the bias arising from variable CpG coverage per gene on the EPIC arrays. Because the harmonized matrix retains only probes shared between EPIC v1 and v2 manifests, CpGs were mapped to EPIC v1 Illumina probe IDs for the enrichment call against a background comprising all CpGs in the harmonized training matrix. Gene assignment followed the default setting, whereby CpGs contribute to a gene if located in any annotated genic region. Terms with a Benjamini-Hochberg FDR < 0.05 were considered significant.

### DMR analysis

DMRs were identified genome-wide from the full cohort using DMRcate [46]. A no-intercept limma model was fitted to test the pairwise diagnostic comparisons IM vs NV, NIM vs NV, and IM vs NIM. Significant DMRs were defined at region-level FDR (HMFDR) < 0.05 and used for overlap analyses with the selected CpGs of each CpG-based model. For downstream interpretation, DMRs with $|\text{mean } \Delta\beta| \geq 0.10$ were considered high-effect DMRs.
To assess whether the NIM methylation profiles fell between NV and IM, per-lesion mean β-values were calculated across the 10,000 CpGs most strongly hypermethylated and the 10,000 most strongly hypomethylated in IM relative to NV across the full cohort.

### Feature importance

To interpret the marker-based and stacked models, permutation feature importance [60] was computed on the external test set for the selected model of each of these two feature sources. Each feature was randomly permuted and the resulting change in the task's primary metric (macro-averaged AUROC for diagnostic classification, macro-averaged MAE for therapeutic group prediction) was recorded, averaged over 100 permutations and oriented so that larger values always denote greater importance (a drop in AUROC, a rise in MAE).

Importance was quantified both per feature and per semantic feature block. For the marker-based models, blocks corresponded to the three data modalities (Horvath residual EAA, EpiSCORE cell type composition, and CNV burden); for the stacked models, blocks corresponded to the two base learners (the CpG-view and marker-view prediction columns). Because the per-class (diagnostic classification) or per-group (therapeutic group prediction) probabilities contributed by a base learner sum to one, permuting them one at a time understates a block's contribution, as the model recovers the signal from the sibling columns; block-level importance therefore permuted all columns of a block jointly under a single shared row order per repeat, preserving each sample's within-block joint distribution while decorrelating the block from the target.

**Marker-target correlation**

To complement the model-based permutation importance with a model-free view, we quantified how each individual methylation-derived marker (Horvath residual EAA, the nine EpiSCORE cell type fractions, and the two CNV burden metrics) tracks the target on its own. The analysis was restricted to the training split. For the diagnostic classification task, the Kruskal-Wallis comparison across the three diagnoses was computed. For the ordinal therapeutic group task, each marker's monotone association was summarized by Spearman's ρ against the therapeutic group alongside a Kruskal-Wallis comparison across groups.

## Statistical Analysis

The primary endpoint for the diagnostic classification task was one vs. rest macro-averaged AUROC, with balanced accuracy and macro-averaged F1 score as secondary endpoints. For the therapeutic group prediction task, macro-averaged, rank based MAE was the primary endpoint, with quadratic weighted Cohen's κ and Spearman's correlation ρ as secondary endpoints. Every test set metric is reported as a point estimate with a 95% bootstrap CI on 1,000 repetitions; resampling was conducted on patient-level.

To assess whether models differ in performance, they were compared pairwise on the external test cohort. For diagnostic classification, the macro-averaged one-vs-rest AUROC and, for therapeutic prediction, the macro-averaged MAE were compared between models using a paired, patient-clustered bootstrap of the difference (1,000 resamples). Each class' one-vs-rest AUROC was compared using DeLong's test [61].

Marker-target associations were examined descriptively on the training cohort only, leaving the test set untouched. For therapeutic prediction, each marker's monotone association with the ordinal group was quantified by Spearman's ρ; for both tasks, differences across diagnosis or therapeutic groups were tested for significance with the Kruskal-Wallis H test [62], and, when the omnibus test was significant, followed by post-hoc two-sided Mann-Whitney U tests for independent samples [63,64] between each pair of groups. Spearman's ρ was omitted for diagnostic classification, whose label is nominal. Throughout, families of related p-values (the per-class AUROC comparisons, the across-marker omnibus tests, and each set of post-hoc pairwise contrasts) were corrected for multiple comparisons with the Holm-Bonferroni procedure [65]. All tests were two-sided with $\alpha = 0.05$.

All statistical analyses were performed using Python 3.14 (*SciPy* 1.17.1, *statsmodels* 0.14.6, *MLstatkit* 0.1.91, *scikit-learn* 1.8.0, *NumPy* 2.4.4) and R 4.6.1 with *Bioconductor* 3.23 (*DMRcate* 3.7.0, *limma* 3.68.0, *missMethyl* 1.45.0).

## Data Availability

External research projects may request access to the prospectively and consecutively collected dataset utilized in our study, specifically for the purpose of advancing skin (cancer) research. Access is granted following an application and approval process managed by the SCP Data Protection Committee, which evaluates requests based on criteria such as alignment with patient consent, a valid ethics vote, and other relevant requirements (i.e., non-commercial [skin] cancer research). All remaining data are available in the article, Supplementary Information, and Supplementary Data files. Commercial use of the dataset is strictly prohibited.

The numerical data underlying all figures and the code used to generate the figures are provided in the GitHub repository (see the Code Availability statement).

## Code Availability

All code is available via GitHub at https://github.com/DBO-DKFZ/methylation_profiling.

# Acknowledgement

We thank the Microarray Core Facility of the DKFZ for providing the Illumina methylation arrays and related services.

# Author Contributions

Jana T. Winterstein: Writing of Original Draft, Conceptualization, Study Design, Analysis, Methodology, Data Acquisition and Curation, Visualization

Lukas Heinlein: Writing of Original Draft, Conceptualization, Study Design, Analysis, Methodology, Software, Visualization

Günter Raddatz: Conceptualization, Analysis, Methodology, Review & Editing

Carina Nogueira Garcia: Conceptualization, Review & Editing

Sarah Haggenmüller: Conceptualization, Data Acquisition and Curation, Review & Editing

Christoph Wies: Conceptualization, Statistical Analysis, Review & Editing

Lucas Schneider: Data Acquisition and Curation, Review & Editing

Annemarie Hoffsommer: Review & Editing
Tim Zeuner: Review & Editing
Friedegund Meier: Resources, Review & Editing
Sarah Hobelsberger: Resources, Review & Editing
Frank F. Gellrich: Resources, Review & Editing
Mildred Sergon: Resources, Review & Editing
Axel Hauschild: Resources, Review & Editing
Lucie Heinzerling: Resources, Review & Editing
Justin G. Schlager: Resources, Review & Editing
Kamran Ghoreschi: Resources, Review & Editing
Max Schlaak: Resources, Review & Editing
Franz J. Hilke: Resources, Review & Editing
Carola Berking: Resources, Review & Editing
Markus V. Heppt: Resources, Review & Editing
Michael Erdmann: Resources, Review & Editing
Sebastian Haferkamp: Resources, Review & Editing
Konstantin Drexler: Resources, Review & Editing
Dirk Schadendorf: Resources, Review & Editing
Wiebke Sondermann: Resources, Review & Editing
Matthias Goebeler: Resources, Review & Editing
Bastian Schilling: Resources, Review & Editing
Daniel Lipka: Resources, Review & Editing
Stefan Fröhling: Resources, Review & Editing
Jakob N. Kather: Conceptualization, Review & Editing
Felix Sahm: Resources, Review & Editing
Yuri Tolkach: Conceptualization, Review & Editing
Jochen S. Utikal: Conceptualization, Resources, Review & Editing
Benjamin Izar: Resources, Review & Editing
Yevgeniy R. Semenov: Resources, Review & Editing
Titus J. Brinker: Supervision, Conceptualization, Review & Editing, Project Administration, Funding Acquisition
Jana T. Winterstein and Lukas Heinlein contributed equally to this work.

# Conflict of Interest Statement

Sarah Haggenmüller reports holding a position at HEINE Optotechnik GmbH & Co. KG. Friedegund Meier reported grants from Novartis and Roche; other (travel support or/and speaker's fees or/and advisor's honoraria) from BMS, MSD, and Pierre Fabre outside the submitted work. Sarah Hobelsberger reported clinical trial support from Almirall and Pierre Fabre, advisor's honoraria from Almirall and Bristol Myers Squibb, speaker's honoraria from Almirall, UCB, Bristol Myers Squibb, Regeneron and AbbVie and travel support from UCB, Janssen Cilag, Almirall, Novartis, Lilly, LEO Pharma, Pierre Fabre and AbbVie outside the submitted work. Axel Hauschild reports personal fees for advisory board membership from Agenus, BMS, Dermagnostix, Eisai, Highlight Therapeutics, Immunocore, Incyte, IO Biotech, Merck Pfizer, Moderna, MSD, Neracare, Novartis, Philogen, Pierre Fabre, Regeneron, Replimune, Roche, Sanofi, Seagen, Skyline Dx and Xenthera. Lucie Heinzerling reported other (clinical studies) from BMS, MSD, Pierre Fabre, Replimune, and Sanofi; personal fees from Biomedx, BMS, MSD, Sun Pharma, Pierre Fabre, Novartis, and Sanofi; and grants from Therakos outside the submitted work. Konstantin Drexler reported speaker's fees and advisor's honoraria from BMS, MSD, Novartis, Sun Pharma, Regeneron and Pierre Fabre outside the submitted work. Sebastian Haferkamp reported speaker's fees and advisor's honoraria from BMS, MSD, Novartis and Pierre Fabre outside the submitted work. Max Schlaak reported personal fees from BMS, Novartis, Immunocore, Kyowa Kirin, Sun Pharma, MSD, Pierre Fabre, outside the submitted work. Carola Berking reported personal fees from BMS, MSD, InflaRx, Novartis, Sanofi, Almirall Hermal, Pierre Fabre, Immunocore, Regeneron, SkylineDx, and Delcath outside the submitted work. Markus V. Heppt received honoraria from Sanofi, Regeneron, Almirall, Biofrontera, Galderma, Novartis, BMS, MSD, Roche, Immunocore, Infectopharm, Pierre Fabre. Michael Erdmann received travel support and speaker's honoraria from Immunocore, Novartis, Pierre Fabre, and Sanofi outside the submitted work. Wiebke Sondermann reported grants from Almirall, Novartis and Medi GmbH; and personal fees from AbbVie, Almirall, BMS, Boehringer Ingelheim, Celgene, Incyte, Janssen, LEO Pharma, Lilly, Novartis, Pfizer, Sanofi Genzyme, Takeda, and UCB outside the submitted work. Matthias Goebeler reported grants for clinical studies to his institution from Argenx, Galderma, Janssen, Novartis, Sanofi, and UCB; personal fees from Almirall (consulting, speaker), Biotest (advisory board), Fresenius-Kabi (advisory board) Janssen (advisory board, speaker), GSK (advisory board, speaker), Lilly (speaker) and Novartis (speaker) and travel support from Abbvie and Janssen outside the submitted work. Bastian Schilling reported speaker's fees and advisor's honoraria from SUN Pharma, Immunocore, Bristol-Myers Squibb, Sanofi, Regeneron, Pierre Fabre Pharma, and Blueprint Medicines, outside the submitted work. Gabriela Poch received travel support and speaker's honoraria from MSD, BMS, Novartis, Sun Pharma and Amgen outside the submitted work. Daniel Lipka contributed to patent filings related to the diagnosis, classification and therapy of hematopoietic neoplasms that are held by the German Cancer Research Center and reports honoraria from Infectopharm GmbH.

Stefan Fröhling reports research funding from Oxford Nanopore Technologies, outside the submitted work. Felix Sahm reports to be founder and advisor of Heidelberg Epignostix GmbH, outside the submitted work. Jakob N. Kather holds shares in StratifAI, Synagen, Spira Labs, Tremont AI, and Saterra AI; is Co-PI on institutional research grants from GSK and AstraZeneca, and declares honoraria or consulting fees from AstraZeneca, Bayer, Bioptimus, Daiichi Sankyo, Eisai, Janssen, Merck, MSD, Novartis, BMS, Roche, and Pfizer. Jochen S. Utikal reported personal fees from Amgen, Bristol Myers Squibb, GSK, Immunocore, LEO Pharma, Merck Sharp & Dohme, Novartis, Pierre Fabre, Rheacell, Roche, and Sanofi outside the submitted work. Benjamin Izar is a consultant for or received honoraria from Volastra Therapeutics, Johnson & Johnson/Janssen, Novartis, GSK, Eisai, AstraZeneca and Merck and has received research funding to Columbia University from Agenus, Alkermes, Arcus Biosciences, Checkmate Pharmaceuticals, Compugen, Immunocore, Regeneron and Synthekine and is cofounder of Basima Therapeutics, outside the submitted work. Yevgeniy R. Semenov is an advisory board member or consultant and has received honoraria from Arcutis, Alterome, Incyte Corporation, Iovance Biotherapeutics, Galderma, Pfizer, Regeneron, and Sanofi outside of the scope of the submitted work. Titus J. Brinker received honoraria from Novartis, Roche, and HEINE Optotechnik, and reported being owner of Smart Health Heidelberg GmbH which released the first two teledermatology services in Germany (“AppDoc” and “Intimarzt”). No other disclosures were reported by any of the authors.

# Literature

[1] Liyanage VRB, Jarmasz JS, Murugeshan N, Del Bigio MR, Rastegar M, Davie JR. DNA Modifications: Function and Applications in Normal and Disease States. Biology 2014;3:670–723. https://doi.org/10.3390/biology3040670.

[2] Loyfer N, Magenheim J, Peretz A, Cann G, Bredno J, Klochendler A, et al. A DNA methylation atlas of normal human cell types. Nature 2023;613:355–64. https://doi.org/10.1038/s41586-022-05580-6.

[3] Anichini A, Caruso FP, Lagano V, Noviello TMR, Tufano R, Nicolini G, et al. Integrated multi-omics profiling reveals the role of the DNA methylation landscape in shaping biological heterogeneity and clinical behaviour of metastatic melanoma. J Exp Clin Cancer Res 2025;44:212. https://doi.org/10.1186/s13046-025-03474-9.

[4] Dor Y, Cedar H. Principles of DNA methylation and their implications for biology and medicine. The Lancet 2018;392:777–86. https://doi.org/10.1016/S0140-6736(18)31268-6.

[5] Wouters J, Vizoso M, Martinez-Cardus A, Carmona FJ, Govaere O, Laguna T, et al. Comprehensive DNA methylation study identifies novel progression-related and prognostic markers for cutaneous melanoma. BMC Med 2017;15:101. https://doi.org/10.1186/s12916-017-0851-3.

[6] Schadendorf D, van Akkooi ACJ, Berking C, Griewank KG, Gutzmer R, Hauschild A, et al. Melanoma. The Lancet 2018;392:971–84. https://doi.org/10.1016/S0140-6736(18)31559-9.

[7] Keung EZ, Gershenwald JE. The eighth edition American Joint Committee on Cancer (AJCC) melanoma staging system: implications for melanoma treatment and care. Expert Rev Anticancer Ther 2018;18:775–84. https://doi.org/10.1080/14737140.2018.1489246.

[8] Haggenmüller S, Wies C, Abels J, Winterstein JT, Heinlein L, Nogueira Garcia C, et al. Discordance, accuracy and reproducibility study of pathologists' diagnosis of melanoma and melanocytic tumors. Nat Commun 2025;16:789. https://doi.org/10.1038/s41467-025-56160-x.

[9] Brochez L, Verhaeghe E, Grosshans E, Haneke E, Piérard G, Ruiter D, et al. Inter-observer variation in the histopathological diagnosis of clinically suspicious pigmented skin lesions. J Pathol 2002;196:459–66. https://doi.org/10.1002/path.1061.
[10] Akbani R, Akdemir KC, Aksoy BA, Albert M, Ally A, Amin SB, et al. Genomic Classification of Cutaneous Melanoma. Cell 2015;161:1681–96. https://doi.org/10.1016/j.cell.2015.05.044.
[11] Cheng L, Lopez-Beltran A, Massari F, MacLennan GT, Montironi R. Molecular testing for BRAF mutations to inform melanoma treatment decisions: a move toward precision medicine. Mod Pathol 2018;31:24–38. https://doi.org/10.1038/modpathol.2017.104.
[12] Ohsie SJ, Sarantopoulos GP, Cochran AJ, Binder SW. Immunohistochemical characteristics of melanoma. J Cutan Pathol 2008;35:433–44. https://doi.org/10.1111/j.1600-0560.2007.00891.x.
[13] Hovestadt V, Remke M, Kool M, Pietsch T, Northcott PA, Fischer R, et al. Robust molecular subgrouping and copy-number profiling of medulloblastoma from small amounts of archival tumour material using high-density DNA methylation arrays. Acta Neuropathol (Berl) 2013;125:913–6. https://doi.org/10.1007/s00401-013-1126-5.
[14] Jurmeister P, Glöß S, Roller R, Leitheiser M, Schmid S, Mochmann LH, et al. DNA methylation-based classification of sinonasal tumors. Nat Commun 2022;13:7148. https://doi.org/10.1038/s41467-022-34815-3.
[15] Capper D, Jones DTW, Sill M, Hovestadt V, Schrimpf D, Sturm D, et al. DNA methylation-based classification of central nervous system tumours. Nature 2018;555:469–74. https://doi.org/10.1038/nature26000.
[16] Koelsche C, Schrimpf D, Stichel D, Sill M, Sahm F, Reuss DE, et al. Sarcoma classification by DNA methylation profiling. Nat Commun 2021;12:498. https://doi.org/10.1038/s41467-020-20603-4.
[17] Conway K, Edmiston SN, Parker JS, Kuan PF, Tsai Y-H, Groben PA, et al. Identification of a robust methylation classifier for cutaneous melanoma diagnosis. J Invest Dermatol 2019;139:1349–61. https://doi.org/10.1016/j.jid.2018.11.024.
[18] Daenekas B, Pérez E, Boniolo F, Stefan S, Benfatto S, Sill M, et al. Conumee 2.0: enhanced copy-number variation analysis from DNA methylation arrays for humans and mice. Bioinformatics 2024;40:btae029. https://doi.org/10.1093/bioinformatics/btae029.
[19] Teschendorff AE, Zhu T, Breeze CE, Beck S. EPISCORE: cell type deconvolution of bulk tissue DNA methylomes from single-cell RNA-Seq data. Genome Biol 2020;21:221. https://doi.org/10.1186/s13059-020-02126-9.
[20] Horvath S, Oshima J, Martin GM, Lu AT, Quach A, Cohen H, et al. Epigenetic clock for skin and blood cells applied to Hutchinson Gilford Progeria Syndrome and ex vivo studies. Aging 2018;10:1758–75. https://doi.org/10.18632/aging.101508.
[21] Ebbelaar CF, Jansen AML, Bloem LT, Blokx WAM. Genome-wide copy number variations as molecular diagnostic tool for cutaneous intermediate melanocytic lesions: a systematic review and individual patient data meta-analysis. Virchows Arch 2021;479:773–83. https://doi.org/10.1007/s00428-021-03095-5.
[22] Fu Q, Chen N, Ge C, Li R, Li Z, Zeng B, et al. Prognostic value of tumor-infiltrating lymphocytes in melanoma: a systematic review and meta-analysis. Oncoimmunology 2019;8:1593806. https://doi.org/10.1080/2162402X.2019.1593806.
[23] Azimi F, Scolyer RA, Rumcheva P, Moncrieff M, Murali R, McCarthy SW, et al. Tumor-Infiltrating Lymphocyte Grade Is an Independent Predictor of Sentinel Lymph Node Status and Survival in Patients With Cutaneous Melanoma. J Clin Oncol 2012;30:2678–83. https://doi.org/10.1200/JCO.2011.37.8539.
[24] Zheng C, Berger NA, Li L, Xu R. Epigenetic age acceleration and clinical outcomes in gliomas. PLoS ONE 2020;15:e0236045. https://doi.org/10.1371/journal.pone.0236045.
[25] Hong C, Yang S, Wang Q, Zhang S, Wu W, Chen J, et al. Epigenetic Age Acceleration of Stomach Adenocarcinoma Associated With Tumor Stemness Features, Immunoactivation, and Favorable Prognosis. Front Genet 2021;12. https://doi.org/10.3389/fgene.2021.563051.
[26] Fan X, Yuan H, Zhao S, Yang X, Shi R, Wang J, et al. Epigenetic age acceleration of early stage hepatocellular carcinoma tightly associated with hepatitis B virus load, immunoactivation, and improved survival. Cancer Biol Ther n.d.;21:899–906. https://doi.org/10.1080/15384047.2020.1804284.

[27] Haggenmüller S, Heinlein L, Abels J, Winterstein JT, Wies C, Kuehn A, et al. Expert-informed artificial intelligence for melanoma classification using dermoscopic and histopathologic data. Nat Commun 2026;In press.
[28] National Comprehensive Cancer Network. NCCN Clinical Practice Guidelines in Oncology: Melanoma: Cutaneous. National Comprehensive Cancer Network; 2026.
[29] Garbe C, Amaral T, Peris K, Hauschild A, Arenberger P, Basset-Seguin N, et al. European consensus-based interdisciplinary guideline for melanoma. Part 1: Diagnostics - Update 2024. Eur J Cancer 2025;215:115152. https://doi.org/10.1016/j.ejca.2024.115152.
[30] Garbe C, Amaral T, Peris K, Hauschild A, Arenberger P, Basset-Seguin N, et al. European consensus-based interdisciplinary guideline for melanoma. Part 2: Treatment – Update 2024. Eur J Cancer 2025;215:115153. https://doi.org/10.1016/j.ejca.2024.115153.
[31] Shain AH, Yeh I, Kovalyshyn I, Sriharan A, Talevich E, Gagnon A, et al. The Genetic Evolution of Melanoma from Precursor Lesions. N Engl J Med 2015;373:1926–36. https://doi.org/10.1056/NEJMoa1502583.
[32] Hartman E, Scott AM, Karlsson C, Mohanty T, Vaara ST, Linder A, et al. Interpreting biologically informed neural networks for enhanced proteomic biomarker discovery and pathway analysis. Nat Commun 2023;14:5359. https://doi.org/10.1038/s41467-023-41146-4.
[33] Elmarakeby HA, Hwang J, Arafeh R, Crowdis J, Gang S, Liu D, et al. Biologically informed deep neural network for prostate cancer discovery. Nature 2021;598:348–52. https://doi.org/10.1038/s41586-021-03922-4.
[34] Zhang S, Li P, Wang S, Zhu J, Huang Z, Cai F, et al. BioM2: biologically informed multi-stage machine learning for phenotype prediction using omics data. Brief Bioinform 2024;25:bbae384. https://doi.org/10.1093/bib/bbae384.
[35] Chanda T, Wies C, Schramm F, Garcia CN, Merl NB, Hetz MJ, et al. Large language models improve physician accuracy but lead to false reliance 2026. https://doi.org/10.48550/arXiv.2608.00817.
[36] Griewank KG, Koelsche C, van de Nes JAP, Schrimpf D, Gessi M, Möller I, et al. Integrated Genomic Classification of Melanocytic Tumors of the Central Nervous System Using Mutation Analysis, Copy Number Alterations, and DNA Methylation Profiling. Clin Cancer Res 2018;24:4494–504. https://doi.org/10.1158/1078-0432.CCR-18-0763.
[37] Jurmeister P, Wrede N, Hoffmann I, Vollbrecht C, Heim D, Hummel M, et al. Mucosal melanomas of different anatomic sites share a common global DNA methylation profile with cutaneous melanoma but show location-dependent patterns of genetic and epigenetic alterations. J Pathol 2022;256:61–70. https://doi.org/10.1002/path.5808.
[38] Hirsch D, Kemmerling R, Davis S, Camps J, Meltzer PS, Ried T, et al. Chromothripsis and focal copy number alterations determine poor outcome in malignant melanoma. Cancer Res 2013;73:1454–60. https://doi.org/10.1158/0008-5472.CAN-12-0928.
[39] Yin Q, Stevenson-Hoare J, Holleczek B, Ben Schöttker, Brenner H. Epigenetic aging and cancer incidence in a German cohort of older adults. Npj Aging 2026;12:41. https://doi.org/10.1038/s41514-026-00356-y.
[40] Rodríguez-Paredes M, Bormann F, Raddatz G, Gutekunst J, Lucena-Porcel C, Köhler F, et al. Methylation profiling identifies two subclasses of squamous cell carcinoma related to distinct cells of origin. Nat Commun 2018;9:577. https://doi.org/10.1038/s41467-018-03025-1.
[41] Banila C, Green D, Katsanos D, Viana J, Osmaston A, Menendez Vazquez A, et al. A noninvasive method for whole-genome skin methylome profiling. Br J Dermatol 2023;189:750–9. https://doi.org/10.1093/bjd/ljad316.
[42] Rodríguez-Paredes M, Feng Y, Gilliam O, Wegner K, Raddatz G, Grönniger E, et al. Non-invasive epidermis sampling for DNA methylation-based prediction of skin cancer phenotypes. Npj Precis Oncol 2026;10:89. https://doi.org/10.1038/s41698-026-01302-7.
[43] Koelsche C, von Deimling A. Methylation classifiers: Brain tumors, sarcomas, and what's next. Genes Chromosomes Cancer 2022;61:346–55. https://doi.org/10.1002/gcc.23041.
[44] Heinlein L, Maron RC, Hekler A, Haggenmüller S, Wies C, Utikal JS, et al. Prospective multicenter study using artificial intelligence to improve dermoscopic melanoma diagnosis in patient care. Commun Med 2024;4:177. https://doi.org/10.1038/s43856-024-00598-5.

[45] Zhou W, Triche TJ Jr, Laird PW, Shen H. SeSAMe: reducing artifactual detection of DNA methylation by Infinium BeadChips in genomic deletions. Nucleic Acids Res 2018;46:e123. https://doi.org/10.1093/nar/gky691.
[46] Peters TJ, Buckley MJ, Statham AL, Pidsley R, Samaras K, V Lord R, et al. De novo identification of differentially methylated regions in the human genome. Epigenetics Chromatin 2015;8:6. https://doi.org/10.1186/1756-8935-8-6.
[47] Zhou W, Laird PW, Shen H. Comprehensive characterization, annotation and innovative use of Infinium DNA methylation BeadChip probes. Nucleic Acids Res 2017;45:e22. https://doi.org/10.1093/nar/gkw967.
[48] Plaksienko A, Di Lena P, Nardini C, Angelini C. methyLImp2: faster missing value estimation for DNA methylation data. Bioinformatics 2024;40:btae001. https://doi.org/10.1093/bioinformatics/btae001.
[49] Pelegí-Sisó D, de Prado P, Ronkainen J, Bustamante M, González JR. methylclock: a Bioconductor package to estimate DNA methylation age. Bioinformatics 2021;37:1759–60. https://doi.org/10.1093/bioinformatics/btaa825.
[50] Ganier C, Mazin P, Herrera-Oropeza G, Du-Harpur X, Blakeley M, Gabriel J, et al. Multiscale spatial mapping of cell populations across anatomical sites in healthy human skin and basal cell carcinoma. Proc Natl Acad Sci U S A 2024;121:e2313326120. https://doi.org/10.1073/pnas.2313326120.
[51] Zhu T, Liu J, Beck S, Pan S, Capper D, Lechner M, et al. A pan-tissue DNA methylation atlas enables in silico decomposition of human tissue methylomes at cell-type resolution. Nat Methods 2022;19:296–306. https://doi.org/10.1038/s41592-022-01412-7.
[52] Chawla NV, Bowyer KW, Hall LO, Kegelmeyer WP. SMOTE: Synthetic Minority Over-sampling Technique. J Artif Intell Res 2002;16:321–57. https://doi.org/10.1613/jair.953.
[53] Ernst J, Kellis M. ChromHMM: automating chromatin-state discovery and characterization. Nat Methods 2012;9:215–6. https://doi.org/10.1038/nmeth.1906.
[54] Kundaje A, Meuleman W, Ernst J, Bilenky M, Yen A, Heravi-Moussavi A, et al. Integrative analysis of 111 reference human epigenomes. Nature 2015;518:317–30. https://doi.org/10.1038/nature14248.
[55] Pedregosa F, Bach F, Gramfort A. On the Consistency of Ordinal Regression Methods. J Mach Learn Res 2017;18:1–35.
[56] Sharabiani MTA, Bottle A, Mahani AS. OGBoost: A Python Package for Ordinal Gradient Boosting. arXivOrg 2025. https://arxiv.org/abs/2502.13456v1 (accessed May 22, 2026).
[57] Cao W, Mirjalili V, Raschka S. Rank consistent ordinal regression for neural networks with application to age estimation. Pattern Recognit Lett 2020;140:325–31. https://doi.org/10.1016/j.patrec.2020.11.008.
[58] Shi X, Cao W, Raschka S. Deep neural networks for rank-consistent ordinal regression based on conditional probabilities. Pattern Anal Appl 2023;26:941–55. https://doi.org/10.1007/s10044-023-01181-9.
[59] Phipson B, Maksimovic J, Oshlack A. missMethyl: an R package for analyzing data from Illumina's HumanMethylation450 platform. Bioinformatics 2016;32:286–8. https://doi.org/10.1093/bioinformatics/btv560.
[60] Fisher A, Rudin C, Dominici F. All Models are Wrong, but Many are Useful: Learning a Variable's Importance by Studying an Entire Class of Prediction Models Simultaneously. J Mach Learn Res 2019;20:1–81.
[61] DeLong ER, DeLong DM, Clarke-Pearson DL. Comparing the Areas under Two or More Correlated Receiver Operating Characteristic Curves: A Nonparametric Approach. Biometrics 1988;44:837–45. https://doi.org/10.2307/2531595.
[62] Kruskal WH, Wallis WA. Use of Ranks in One-Criterion Variance Analysis. J Am Stat Assoc 1952;47:583–621. https://doi.org/10.1080/01621459.1952.10483441.
[63] Wilcoxon F. Individual Comparisons by Ranking Methods. Biom Bull 1945;1:80–3. https://doi.org/10.2307/3001968.
[64] Mann HB, Whitney DR. On a Test of Whether one of Two Random Variables is Stochastically Larger than the Other. Ann Math Stat 1947;18:50–60.
[65] Holm S. A Simple Sequentially Rejective Multiple Test Procedure. Scand J Stat 1979;6:65–70.

Supplements to **From Lesion Classification to Therapeutic Stratification: DNA Methylation Profiling in Melanoma**

**Supplementary Table 1. Binning of AJCC 8th edition stages into therapeutic groups**

| AJCC 8th Edition Stage | Therapeutic Group |
|---|---|
| 0 | 0 |
| IA | IA |
| IB | IB |
| IIA | IIA |
| IIB-C | IIB/C |
| IIIA-C | III |
| IV | IV (dropped due to sample size) |

**Supplementary Table 2. Characteristics of the datasets used for development and testing of the diagnostic classification models.** IM: invasive melanoma, NIM: non-invasive melanoma, NV: melanocytic nevus.

| Hospital | Overall (%) | IMs | NIMs | NVs | Purpose |
|---|---|---|---|---|---|
| Hospital 1 | 365 (36.5%) | 141 (40.4%) | 49 (41.9%) | 175 (32.7%) | training |
| Hospital 2 | 128 (12.8%) | 39 (11.2%) | 19 (16.2%) | 70 (13.1%) | training |
| Hospital 3 | 88 (8.8%) | 13 (3.7%) | 10 (8.5%) | 65 (12.1%) | training |
| Hospital 4 | 77 (7.7%) | 47 (13.5%) | 7 (6.0%) | 23 (4.3%) | training |
| Hospital 5 | 64 (6.4%) | 25 (7.2%) | 6 (5.1%) | 33 (6.2%) | training |
| Hospital 6 | 63 (6.3%) | 19 (5.4%) | 6 (5.1%) | 38 (7.1%) | training |
| Hospital 7 | 29 (2.9%) | 18 (5.2%) | 4 (3.4%) | 7 (1.3%) | training |
| Hospital 8 | 187 (18.7%) | 47 (13.5%) | 16 (13.7%) | 124 (23.2%) | testing |
| **Overall** | 1,001 (100.0%) | 349 (100.0%) | 117 (100.0%) | 535 (100.0%) | |

**Supplementary Table 3. Characteristics of the datasets used for development and testing of the therapeutic group prediction models.**

| Hospital | Overall (%) | 0 | IA | IB | IIA | IIB/C | III | Purpose |
|---|---|---|---|---|---|---|---|---|
| Hospital 1 | 186 (43.2%) | 49 (41.9%) | 64 (41.6%) | 34 (43.6%) | 7 (33.3%) | 19 (52.8%) | 13 (52.0%) | training |
| Hospital 2 | 56 (13.0%) | 19 (16.2%) | 17 (11.0%) | 13 (16.7%) | 1 (4.8%) | 3 (8.3%) | 3 (12.0%) | training |
| Hospital 3 | 16 (3.7%) | 10 (8.5%) | 4 (2.6%) | 2 (2.6%) | 0 (0.0%) | 0 (0.0%) | 0 (0.0%) | training |
| Hospital 4 | 44 (10.2%) | 7 (6.0%) | 22 (14.3%) | 10 (12.8%) | 1 (4.8%) | 2 (5.6%) | 2 (8.0%) | training |

| Hospital 5 | 27 (6.3%) | 6 (5.1%) | 9 (5.8%) | 6 (7.7%) | 4 (19.0%) | 2 (5.6%) | 0 (0.0%) | training |
|---|---|---|---|---|---|---|---|---|
| Hospital 6 | 23 (5.3%) | 6 (5.1%) | 11 (7.1%) | 2 (2.6%) | 2 (9.5%) | 2 (5.6%) | 0 (0.0%) | training |
| Hospital 7 | 21 (4.9%) | 4 (3.4%) | 8 (5.2%) | 5 (6.4%) | 0 (0.0%) | 4 (11.1%) | 0 (0.0%) | training |
| Hospital 8 | 58 (13.5%) | 16 (13.7%) | 19 (12.3%) | 6 (7.7%) | 6 (28.6%) | 4 (11.1%) | 7 (28.0%) | testing |
| **Overall** | 431 (100.0%) | 117 (100.0%) | 154 (100.0%) | 78 (100.0%) | 21 (100.0%) | 36 (100.0%) | 25 (100.0%) | |

**Supplementary Table 4. Patient characteristics.** std: standard deviation, IQR: interquartile range.

| **Patient characteristic** | | **n = 923** | **%** |
|---|---|---|---|
| Age | [mean (std, IQR)] | 58.6 | (18.8, 31.0) |
| | <35 | 128 | 13.9 |
| | 35-54 | 239 | 25.9 |
| | 55-74 | 329 | 35.6 |
| | >74 | 227 | 24.6 |
| Sex | | | |
| | Male | 522 | 56.6 |
| | Female | 401 | 43.5 |
| Fitzpatrick skin phototype | | | |
| | I | 80 | 8.7 |
| | II | 574 | 62.2 |
| | III | 232 | 25.1 |
| | IV | 19 | 2.1 |
| | V | 2 | 0.2 |
| | VI | 2 | 0.2 |
| | Unknown | 14 | 1.5 |

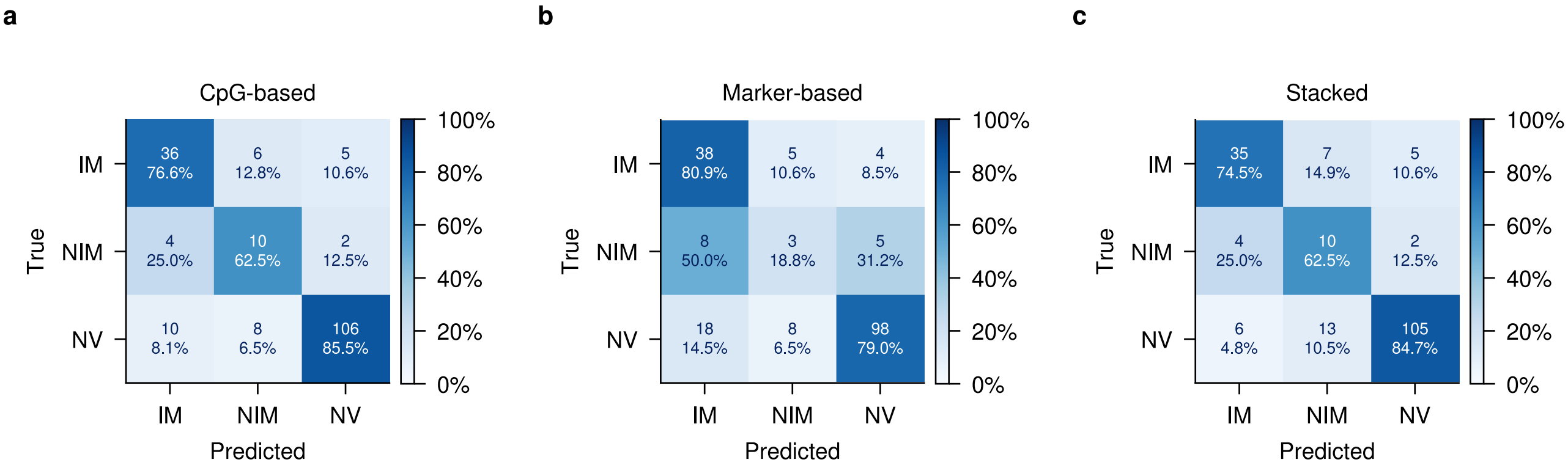

**Supplementary Figure 1. Confusion Matrices for the best-performing diagnostic classification models.**

Results are displayed for the model based on **a** CpGs, **b** methylation-derived markers, and **c** stacked. IM: invasive melanoma, NIM: non-invasive melanoma, NV: melanocytic nevus, SVM: support vector machine, RF: Random Forest, MLP: Multi-layer perceptron.

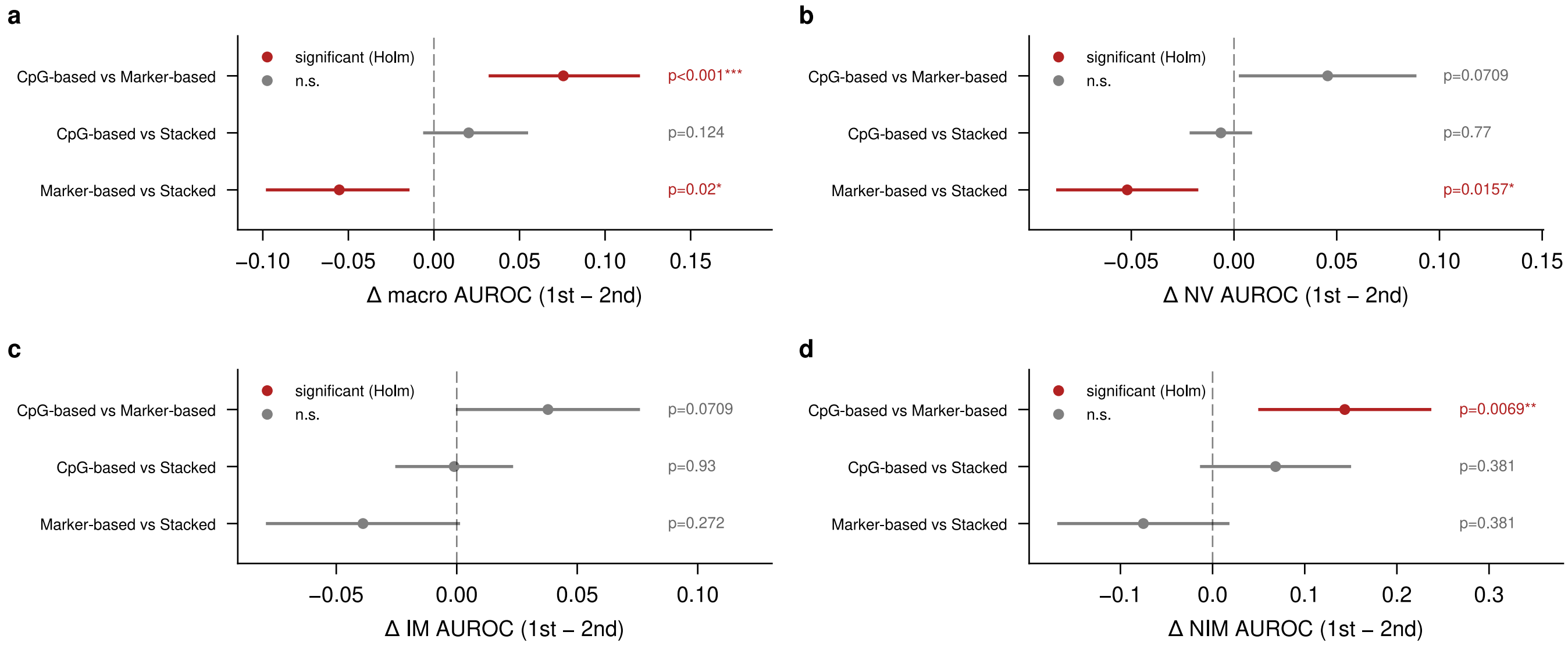

**Supplementary Figure 2. Pairwise comparison of the area under the receiver operator characteristic curve (AUROC) for the best-performing diagnostic classification models.**

**a** paired, patient-clustered bootstrap of the difference (1,000 resamples) in macro-averaged AUROC. DeLong tests for **b** melanocytic nevus (NV) one-vs-rest AUROC **c** invasive melanoma (IM) one-vs-rest AUROC **d** non-invasive melanoma (NIM) one-vs-rest AUROC. ***: $p < 0.001$, **: $p < 0.01$, *: $p < 0.05$.

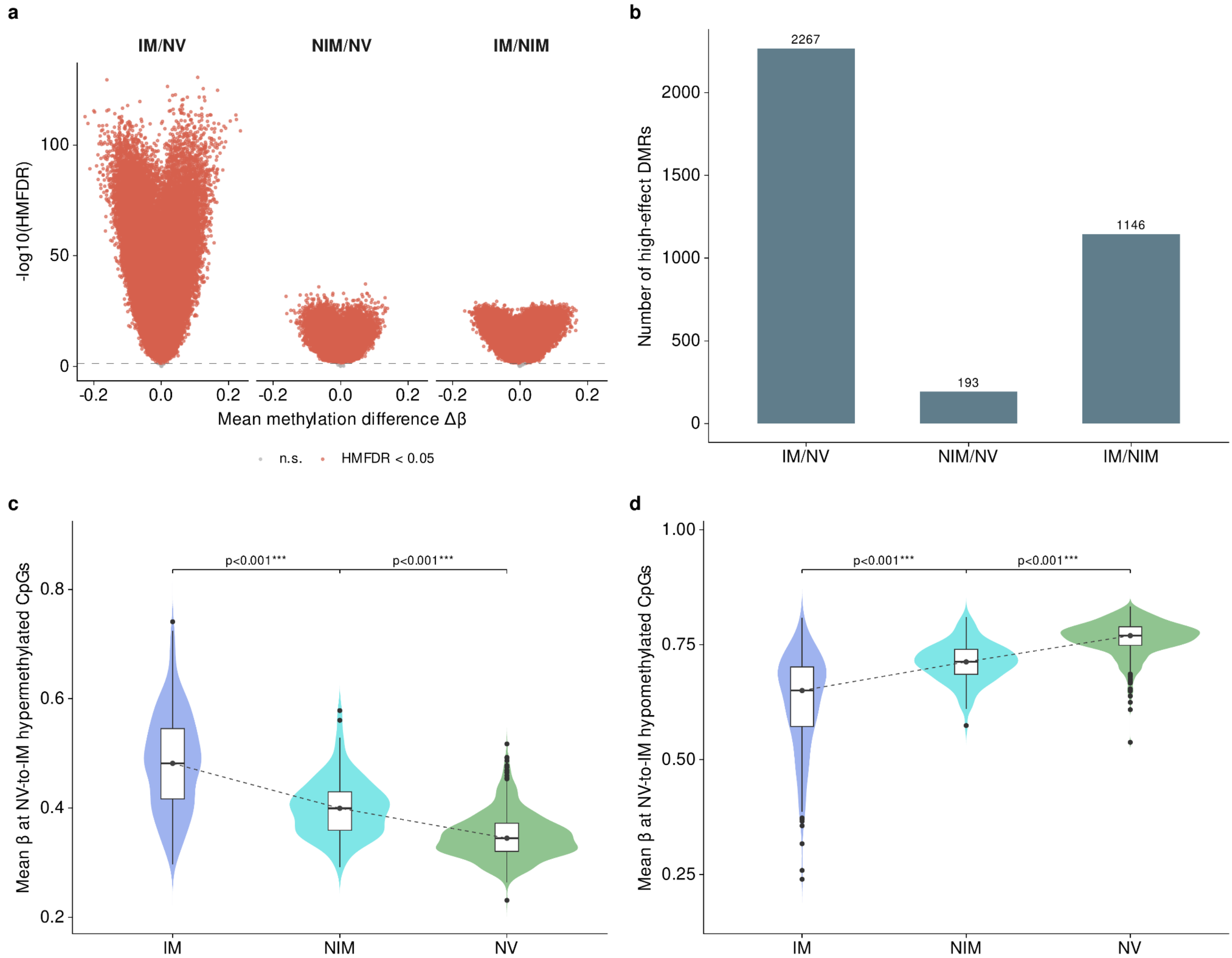

**Supplementary Figure 3. DNA methylation across melanocytic nevi, non-invasive melanoma and invasive melanoma. a** Volcano plots of genome-wide DMRs across pairwise diagnostic comparisons. Dashed lines indicate a HMFDR threshold of 0.05; significant DMRs above the threshold are coloured orange. **b** Number of significant high-effect DMRs per pairwise diagnostic comparison. **c** and **d** Per-lesion mean β-values across the 10,000 most strongly hypermethylated (**c**) and hypomethylated (**d**) CpGs between melanocytic nevi (NV) and invasive melanoma (IM). Violin plots show the distribution across lesions, with box plots indicating the median and interquartile range (IQR). Dashed lines connect group medians. IM vs non-invasive melanoma (NIM) and NIM vs NV were compared by two-sided Wilcoxon rank-sum tests (Holm-adjusted). ***: $p < 0.001$.

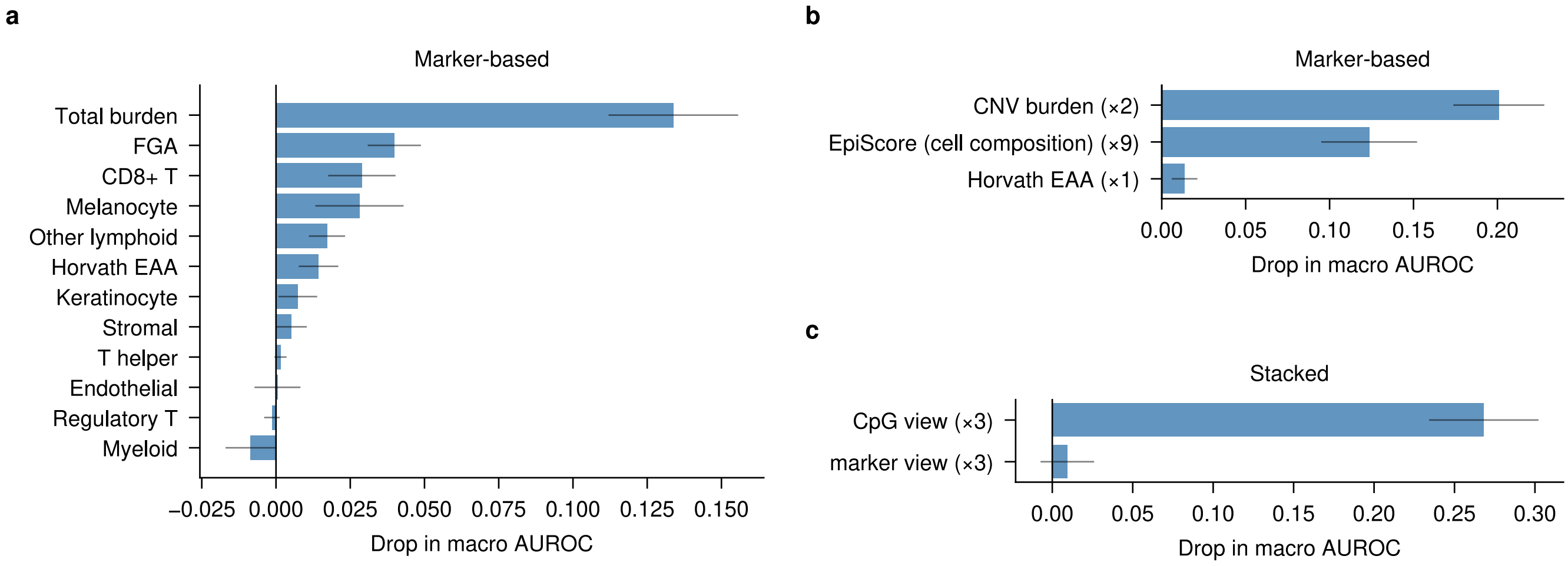

**Supplementary Figure 4. Feature importance of marker and stacked model for the best-performing classification models.**
**a** all features of the marker model **b** the grouped features of the marker model, and **c** the grouped features of the stacked model.

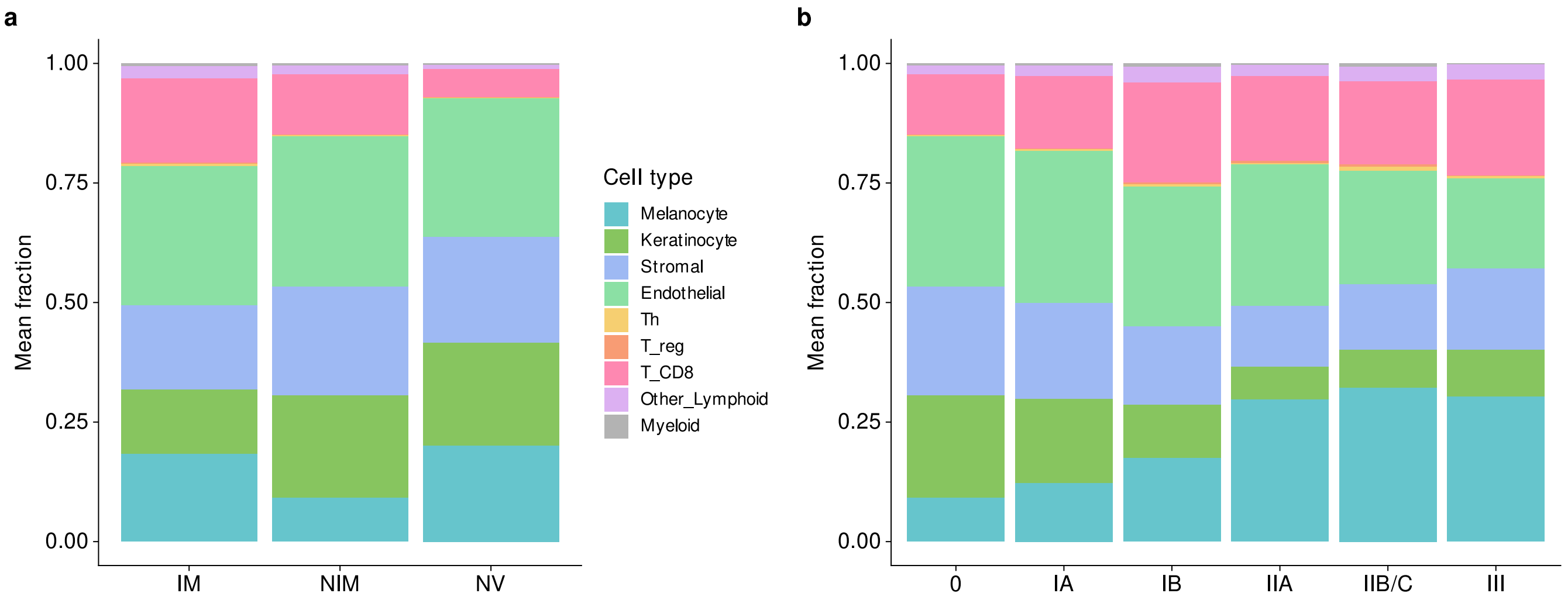

**Supplementary Figure 5. Methylation-based cell type deconvolution.**
Estimated cell type composition is shown per **a** diagnosis **b** therapeutic group. IM: invasive melanoma, NIM: non-invasive melanoma, NV: melanocytic nevus.

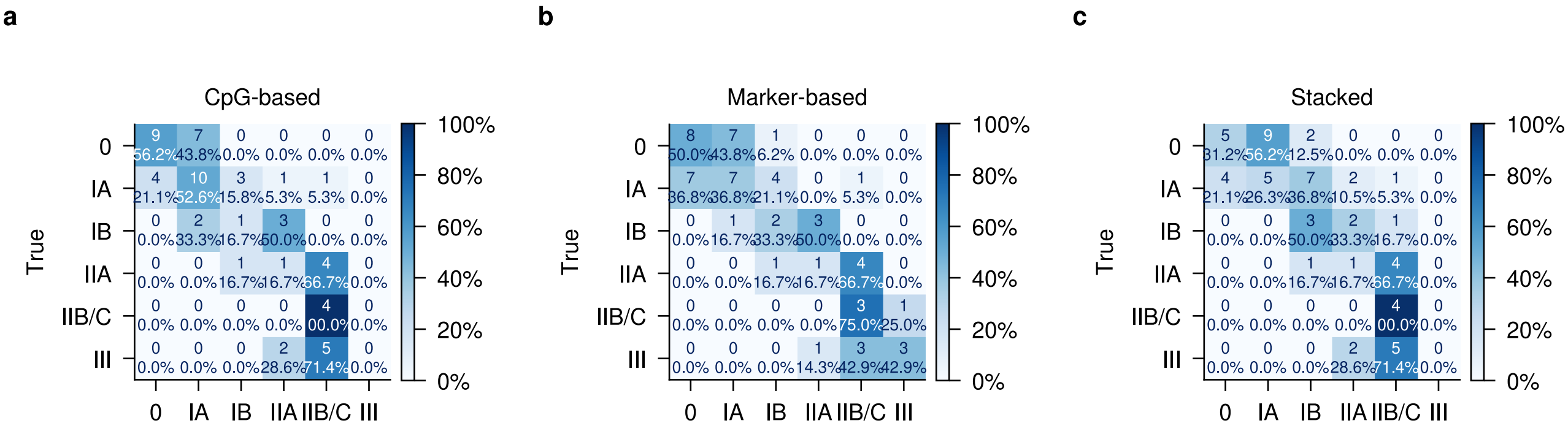

**Supplementary Figure 6. Confusion Matrices for the best-performing therapeutic group prediction models.**

Results are displayed for the model based on **a** CpGs**, b** methylation-derived markers, and **c** stacked.

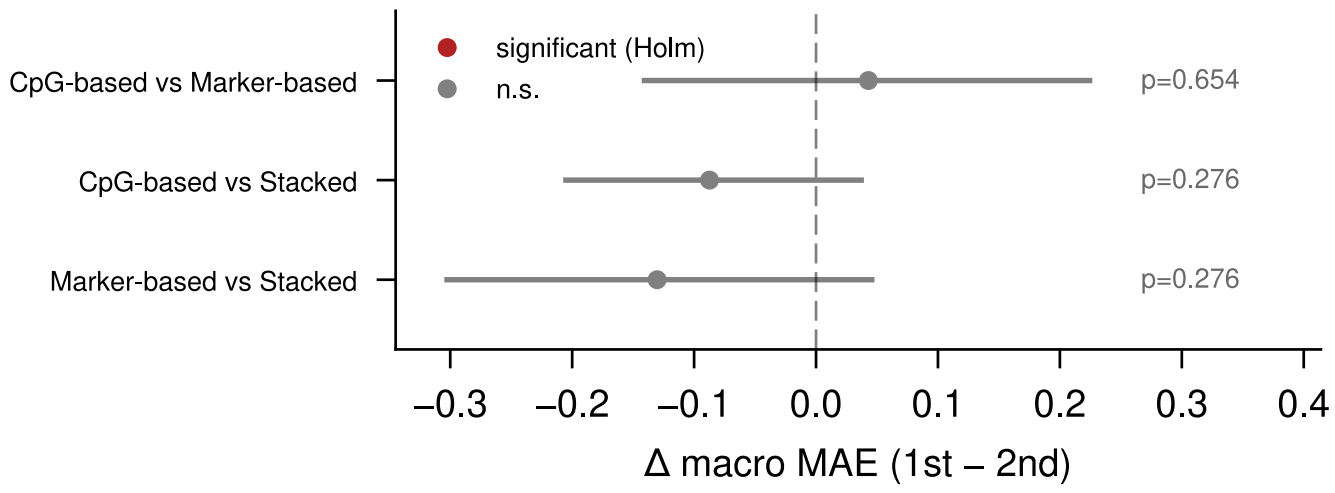

**Supplementary Figure 7. Pairwise comparison of the macro-averaged mean absolute error (MAE) for the best-performing ordinal models.**

Paired, patient-clustered bootstrap of the difference (1,000 resamples).